\documentclass[a4paper,11pt]{article}
\pdfoutput=1 % if your are submitting a pdflatex (i.e. if you have
\usepackage{jheppub}
\usepackage{color}
\usepackage{graphicx}
\usepackage{amsmath}
\usepackage{amssymb}
\usepackage{ulem}
\title{}
\date{}
\renewcommand{\vec}[1]{\mbox{\boldmath$ #1 $}}

\newcommand{\pt}{p_T}
\newcommand{\ptm}{\not\!{p}_T}
\newcommand{\bav}{b_{V1}}
\newcommand{\bbv}{b_{V2}}
\newcommand{\baw}{b_{W1}}
\newcommand{\bbw}{b_{W2}}
\newcommand{\baz}{b_{Z1}}
\newcommand{\bbz}{b_{Z2}}
\newcommand{\ttbar}{t\bar{t}}
\newcommand{\invfb}{{\rm fb}^{-1}}
\newcommand{\dhvv}{h\rightarrow V^{(*)}V^{(*)}}
\newcommand{\tref}[1]{table~\ref{#1}}
\newcommand{\sref}[1]{section~\ref{#1}}
\newcommand{\fref}[1]{figure~\ref{#1}}
\newcommand{\eref}[1]{eq.~(\ref{#1})}
\newcommand{\apref}[1]{appendix~\ref{#1}}

\def\op{\mathcal{O}}
\def\cw{\cos\theta_{w}}

\def\nn{\nonumber \\}
\newcommand{\cwp}[1]{\cos^{#1}\theta_{w}}

\title{\boldmath Jet substructure and probes of CP violation in Vh production}

\author[a]{R. M. Godbole,}
\author[b]{D. J. Miller,}
\author[a]{K. A. Mohan}
\author[b]{and C. D. White}

\affiliation[a]{Centre for High Energy Physics, Indian Institute of 
Science, Bangalore 560 012, India}
\affiliation[b]{School of Physics and Astronomy, Scottish Universities
Physics Alliance, University of Glasgow, Glasgow G12 8QQ, Scotland, UK}

\emailAdd{rohini@cts.iisc.ernet.in}
\emailAdd{David.J.Miller@glasgow.ac.uk}
\emailAdd{kirtimaan@cts.iisc.ernet.in}
\emailAdd{Christopher.White@glasgow.ac.uk}

\abstract{
We analyse the $hVV$ ($V=W,Z$) vertex in a model independent way
using $Vh$ production. To that end, we consider 
possible corrections to the Standard Model Higgs
Lagrangian, in the form of higher dimensional operators which
parametrise the effects of new physics.
In our analysis, we pay special attention to linear observables that can be used to probe CP violation in the same.
 By considering the associated
production of a Higgs boson with a vector boson ($W$ or $Z$), we use
jet substructure methods to define angular observables which are
sensitive to new physics effects, including an asymmetry which is
linearly sensitive to the presence of CP odd effects. We demonstrate
how to use these observables to place bounds on the presence of higher
dimensional operators, and quantify these statements using a log
likelihood analysis. Our approach allows one to probe separately the $hZZ$ and
$hWW$ vertices, involving arbitrary combinations of BSM operators,
at the Large Hadron Collider.
}

\begin{document}
\maketitle
\flushbottom

\section{Introduction}
\label{sec:intro}
Both before and after the discovery of a new resonance at the Large
Hadron Collider (LHC)~\cite{Aad:2012tfa,Chatrchyan:2012ufa}, much
attention has been focused on how to efficiently determine its spin and
couplings~\cite{Espinosa:2012ir,Ellis:2012rx,Ellis:2012hz,Espinosa:2012vu,Corbett:2012dm,Espinosa:2012im,Azatov:2012bz,Montull:2012ik,Klute:2012pu,Azatov:2012ga,Gao:2010qx,Choi:2002jk,Ellis:2012wg,DeRujula:2010ys,Odagiri:2002nd,Buszello:2002uu,Bredenstein:2006rh,BhupalDev:2007is,DeSanctis:2011yc,Ellis:2012mj,Boughezal:2012tz,Stolarski:2012ps,Ellis:2012xd,Djouadi:2013yb,Godbole:2013saa,Ellis:2013ywa,Godbole:2011hw,Muhlleitner:2012jy,Boos:2013mqa,Sun:2013yra,Einhorn:2013tja,Anderson:2013afp,Masso:2012eq,Nhung:2013lpa,Heinemeyer:2013tqa,Delaunay:2013npa,Delaunay:2013pja,Maltoni:2013sma,Belusca-Maito:2014dpa,Gavela:2014vra,Biekoetter:2014jwa}.
Deviations from Standard Model (SM) behaviour would signal the
presence of new physics beyond the Standard Model (BSM), and there are
significant motivations for expecting such deviations to be present at
some level, not the least given that new physics is expected to explain or
clarify the nature of electroweak symmetry breaking. There are two
main approaches for addressing BSM corrections to the Higgs
sector. The first is to postulate the existence of a specific theory,
and analyse how the particle content leads to corrections to SM
observables. This approach must necessarily be used for collider
experiments whose energy exceeds the lowest energy scale associated
with the new physics (e.g. a new particle mass). The second
possibility is to use effective field theory techniques to write down
possible corrections to the SM Lagrangian in the form of additional
operators, which ultimately arise from integrating out the new degrees
of freedom in a particular BSM model. One may systematically classify
these operators according to their mass dimension, such that
higher-dimensional ones are suppressed by increasing powers of the new
physics scale. For a given mass dimension, there is a finite set of possible independent operators. 
By including all of these (in a
chosen basis), one allows for the most general corrections to the
SM. This approach has the benefit of being completely
model-independent, but at the price of being applicable only for
energy scales which are below the lowest new physics scale. This is a
reasonable assumption to make, given that current studies (such as
those referred to above) appear to show only small deviations, if any, from
the Standard Model.

In this paper, we focus on the coupling of the Higgs to vector bosons
$V=W$, $Z$. The operators relevant for these interactions have been
classified
in~\cite{Buchmuller:1985jz,Contino:2013kra,Grzadkowski:2010es}. 
It is important to understand that bounds derived on the $hZZ$ vertex, do not automatically translate to 
bounds on the $hWW$ vertex. For example, as argued in ref.~\cite{Delaunay:2013npa}, violation of custodial symmetry can arise
naturally in new physics models. While higher dimension operators may be constrained from precision tests as well as Higgs rates~\cite{Banerjee:2013apa,Belusca-Maito:2014dpa,Ellis:2014dva}, the constraints depend on various assumptions. Unambiguous and definitive constraints 
can only be determined by directly probing the nature of the $hZZ$ and $hWW$ vertices \textit{separately}.
In
order to determine whether or not the higher dimension operators are present in nature, one must
study various scattering processes that involve the $hZZ$ and $hWW$
vertices. The decay of the Higgs boson to $Z$ boson pairs at the LHC
has been studied
in~\cite{Choi:2002jk,Gao:2010qx,DeRujula:2010ys,Desai:2011yj,Stolarski:2012ps,Bolognesi:2012mm,Chen:2014ona,Chatrchyan:2013iaa,CMS-PAS-HIG-14-014,2013arXiv1310.8361D},
which focused on the fully leptonic decay channel. Combined with LHC
data, this disfavours the possibility that the recently discovered
boson is purely pseudoscalar at $\sim2-3\sigma$
significance~\cite{ATLAS-CONF-2013-013,CMS-PAS-HIG-13-002,Chatrchyan:2012jja,PhysRevLett.110.081803,PhysRevD.89.092007}.
The decay of the Higgs to $W$ boson pairs is more difficult in
principle, due to the limited kinematic resolution inherent in having
missing energy in the final state. This mode has been investigated
in~\cite{Bredenstein:2006rh,DeSanctis:2011yc,Gao:2010qx,Ellis:2012wg,Abazov:2014doa,CMS-PAS-HIG-14-012}. As
ref.~\cite{Ellis:2012wg} in particular makes clear, the kinematic cuts
used to select events in this case may overly diminish the signal for
BSM effects.\\

Another possibility is to study the production of the Higgs boson via
vector boson fusion, and angular observables exist for distinguishing
various BSM
scenarios~\cite{Plehn:2001nj,Hankele:2006ma,Andersen:2010zx,Andersen:2008gc}. However,
a deficiency of this mode is that it is not possible to unambiguously
separate BSM contributions to the $hWW$ and $hZZ$
vertices. Furthermore, ref.~\cite{Djouadi:2013yb} argued that the
momentum dependence associated with an anomalous $hVV$ vertex can have
a dramatic effect on the rapidities of the quarks that emit the vector
bosons, and consequently of the acceptance of the event selection
cuts.\\

Given the above difficulties, the possibility has been explored of
using a future electron-positron collider, such as the proposed
International Linear Collider (ILC) or
equivalent~\cite{Miller:2001bi,Han:2000mi,Biswal:2008tg,Biswal:2009ar,Dutta:2008bh}. The
different BSM corrections have different CP properties, which manifest
themselves in different angular decay products of the Higgs and
associated particles. An $e^+e^-$ collider can explore this in detail
using polarised beams. In addition, the partonic centre of mass energy
is known precisely in such a collider, and one may distinguish
different contributions to the $VVh$ vertices using the fact that they
lead to different power-like growths of associated Higgs production
cross-sections near threshold~\cite{Miller:2001bi}. However, it is
still not possible to unambiguously determine BSM corrections to the
$WWh$ and $ZZh$ vertices separately at such a collider~\footnote{
It is not easy to study the anomalous WWh couplings via the process $e^{+}e^{-}\to \nu \bar{\nu} h$ since there are large irreducible 
backgrounds to this process and a high degree of beam polarization as well as measurements of the polarization of the final states are required. See ref.~\cite{Biswal:2008tg} for details.
}.
Whilst this is possible at the LHeC (a proposed $e^-p$
facility)~\cite{AbelleiraFernandez:2012cc,Biswal:2012mp}, it is clearly advantageous to use the
LHC itself to achieve this.\\

In this paper, we show that one can indeed distinguish the presence of
higher dimensional operators at the LHC, using the associated
production of a Higgs with a vector boson ($Vh$ production), where the
Higgs boson decays to a pair of $b$ quarks. For many years, it was
thought to be impossible to analyse this mode, due to the presence of
large QCD backgrounds. This situation has changed due to the
development of jet substructure techniques, as pioneered
in~\cite{Butterworth:2008iy}. By requiring the Higgs boson to be
boosted, the $b$ quark pair from its decay will be approximately
collinear. One may then distinguish the boosted Higgs signal by
looking for a fat jet, containing two smaller subjets (modulo a
filtering procedure) each of which reconstructs the $b$
mass. Subsequently, a number of approaches for utilising jet
substructure have been
developed~\cite{Ellis:2009su,Ellis:2009me,Krohn:2009th,Soper:2011cr,Soper:2010xk},
together with analytic
understanding~\cite{Dasgupta:2013ihk,Dasgupta:2013via} and
applications in experimental
analyses~\cite{ATLAS:2012am,Aad:2012meb,Aad:2013gja,Chatrchyan:2013rla,Aad:2012raa,Aad:2012dpa,ATLAS:2012dp,ATLAS:2012ds,Chatrchyan:2012ku,Chatrchyan:2012cx,Chatrchyan:2012ypy}. As
the present authors already pointed out in~\cite{Godbole:2013saa},
reconstructing both the Higgs momentum and the associated vector boson
opens up the use of polarisation-related methods for $Vh$ production,
analogous to those used in the $e^+e^-$ studies mentioned above: the
spin state of the associated vector boson is influenced by the
presence of higher dimensional operators in the Higgs sector, so that
angular observables involving the vector boson decay products can be
used to constrain BSM physics. Furthermore, this can be done
separately for the $Zh$ and $Wh$ channels, allowing one to
independently elucidate the nature of the $hZZ$ and $hWW$ vertices.\\

The structure of our paper is as follows. Throughout the remainder of
this introduction, we discuss the framework we are using for higher
dimensional operators in more detail. In section~\ref{c6:sec:mc}, we
describe the details of our simulations and the selection cuts. We
also make a note of higher order effects in both $Wh$ and $Zh$
production and describe kinematic reconstruction issues. In
\sref{c6:sec:sens} we describe the increased sensitivity to non-SM
couplings of the $hVV$ vertex, mentioned earlier. In \sref{c6:sec:ang}
we construct and describe various angular observables that are able to
discriminate the different non-SM couplings of the $hVV$ vertex. In
\sref{c6:sec:ll} we construct likelihoods out of various observables
and estimate the required luminosity for the $14$~TeV LHC to constrain
the anomalous $hVV$ couplings. In \sref{c6:sec:asym} we construct a
CP-odd asymmetry that is linearly sensitive to the CP-odd coupling and
hence to CP violating effects in the $hVV$ vertex. Finally in
section~\ref{c6:sec:summary} we summarize and conclude.

\subsection{Higher Dimensional Operators}
\label{sec:higher}
As already mentioned in the introduction, one may encapsulate the
structure of BSM physics in a model-independent way by adding higher
dimensional operators to the SM Lagrangian. One starts by classifying
all possible higher-dimensional operators that can serve as
corrections to the Standard Model Lagrangian and that are
gauge-invariant, an exercise which was first carried out
in~\cite{Burges:1983zg,Leung:1984ni,Buchmuller:1985jz}. There is a
single dimension five operator which, after electroweak symmetry
breaking, is responsible for neutrino masses and mixings. One is then
motivated to proceed to dimension six operators, of which there are
many - ref.~\cite{Buchmuller:1985jz} lists over a hundred. However,
not all of these are independent, as one may use equations of motion
to relate them. To this end, ref.~\cite{Grzadkowski:2010es} showed
that there are 59 independent
operators. Reference~\cite{Contino:2013kra} expressed these in a basis
more directly suited to Higgs boson physics, and also discussed how
the coefficients of these operators scale differently if electroweak
symmetry breaking is weakly or strongly coupled. Effective operators
for a hypothetical spin one or spin two Higgs boson have been
presented in~\cite{Artoisenet:2013puc}, which also discusses their
implementation in a computational framework inclusive of
next-to-leading order matrix element corrections and parton shower
effects. A pedagogical review of the literature may be found in
section 2 of ref.~\cite{Alloul:2013naa}. \\

Since our objective is to study the Lorentz structure of the $hVV$
vertex for massive gauge bosons, we will only concern ourselves with a
subset of the operators. It is sufficient to consider the following
three operators~\footnote{These correspond to the operators ${\cal
    O}_{\Phi W}$, ${\cal O}_{\Phi\tilde{W}}$ and ${\cal O}'_{DW}$ in
  ref.~\cite{Heinemeyer:2013tqa}.}
\begin{align*}
&\op_{WW}=\frac{g_2^2\  b_{WW}}{4\Lambda^2}\Phi^\dagger\Phi W_{\mu\nu}^{i}W^{i\mu\nu}\ , \quad \quad
\widetilde{\op}_{WW}=\frac{g_2^2\  c_{WW}}{4\Lambda^2}\Phi^\dagger\Phi W_{\mu\nu}^{i}\tilde{W}^{i\mu\nu}\ ,&\nn
&\op_{hW}=\frac{i g_{2}\  b_{ hW}}{\Lambda^2} \left(D^\nu W_{\mu\nu}\right)^k
\left(\Phi^\dag\sigma^k \overleftrightarrow{D}^\mu\Phi\right).&
\end{align*}
Here $\Lambda$ is the scale of new physics and the multiplicative
Wilson coefficients $b_{WW}$, $c_{WW}$ and $b_{ hW}$ parametrize the
relative strengths of these operators. Writing the coupling in the
form $i\Gamma^{\mu\nu}(k_1,k_2) \epsilon_\mu(k_1)
\epsilon^\ast_\nu(k_2)$, where $\{\epsilon_\mu(k_i)\}$ ($i=1,2$) are
the polarization vectors of the two gauge bosons, the $HVV$ vertices
one obtains are
\begin{align}
i\Gamma^{\mu\nu}_{hWW}(k_1,k_2)= & 
i\left(g_2 m_W\right) \bigg[ \eta^{\mu\nu}\left(1+ a_W - \frac{\baw}{m_W^2}\left(k_1 \cdot k_2\right) + \frac{\bbw}{m_W^2}\left(k_1^2 + k_2^2\right)\right) \notag \\ 
& + \frac{\baw}{m_W^2}\ k_1^{\nu}  k_2^{\mu}
- \frac{\bbw}{m_W^2}\left(k_1^{\mu}k_1^{\nu} + k_2^{\mu}k_2^{\nu}\right) 
\notag \\
&+ \frac{c_W}{m_W^2}\ \epsilon^{\mu\nu\rho\sigma}k_{1\rho}k_{2\sigma} \bigg]\ ;
\label{c4:eqn:vertw}
\end{align}
\begin{align}
i\Gamma_{hZZ}^{\mu\nu}(k_1,k_2)=& 
i\left(\frac{g_2 m_Z}{\cw}\right) \bigg[ \eta^{\mu\nu}\left(1+ a_Z - \frac{\baz}{m_Z^2}\left(k_1 \cdot k_2\right) + \frac{\bbz}{m_Z^2}\left(k_1^2 + k_2^2\right)\right) \notag \\ 
& + \frac{\baz}{m_Z^2}\ k_1^{\nu}  k_2^{\mu}
- \frac{\bbz}{m_Z^2}\left(k_1^{\mu}k_1^{\nu} + k_2^{\mu}k_2^{\nu}\right) 
\notag \\
&+ \frac{c_Z}{m_Z^2}\ \epsilon^{\mu\nu\rho\sigma}k_{1\rho}k_{2\sigma} \bigg]\ ,
\label{c4:eqn:vertz}
\end{align}
where we have introduced the rescaled parameters
\begin{gather}
\baw=\frac{2\ m_W^2\  b_{WW}}{\Lambda^2}\ , \quad  \bbw=\frac{2\ m_W^2\ b_{hW}}{\Lambda^2}\ , \nn
c_W=\frac{2\ m_W^2\ c_{WW}}{\Lambda^2}\ ,
\label{paramsW}
\end{gather}
\begin{gather}
\baz=\frac{2\cwp{2}\ m_W^2\ b_{WW}}{\Lambda^2}\ , \quad  \bbz=\frac{2m_W^2 b_{hW}}{\Lambda^2}\ , \nn
c_Z=\frac{2\cwp{2}\ m_W^2\ c_{WW}}{\Lambda^2}\ ,
\label{paramsZ}
\end{gather}
and also rescaled the $\eta^{\mu\nu}$ contribution by a factor $1+a_V$
in each case, to allow for full generality (i.e. sensitivity to
rescalings of the Standard Model contribution).
Note that while the terms in the first two lines of each of
eqs.~(\ref{c4:eqn:vertw}, \ref{c4:eqn:vertz}) are CP--even, the terms
in the third lines are CP--odd. Terms which are not proportional to
$\eta^{\mu\nu}$ may be generated within the SM at higher orders of
perturbation theory, although the resulting couplings are likely to be
very small. Significantly large values of these couplings would be a
signal for BSM physics.  \\

One should note that, since we will always consider this vertex in
processes where the V bosons are connected to external fermions, the
terms $\left(\bbw\left(k_1^{\mu}k_1^{\nu} + k_2^{\mu}k_2^{\nu}\right)
\right)$ and $\left(\bbz\left(k_1^{\mu}k_1^{\nu} + k_2^{\mu}k_2^{\nu}
\right)\right)$ vanish due to current conservation and we will not
consider them any further.  Note that the extra factors of $\cw$ in
($\baz$, $c_Z$) as compared to ($\baw$, $c_W$) signal violation of
custodial symmetry. These factors disappear when the corresponding
$\op_{BB}$, $\tilde{\op}_{BB}$ and $\op_{hB}$ operators are of equal
strength.  \\

The aim of this paper is to perform a detailed study of angular
observables designed to distinguish the three contributions to each
$hVV$ vertex: SM, BSM CP even, and BSM CP odd, building upon the
preliminary study of~\cite{Godbole:2013saa}. In what follows we will
present results in terms of the vertex parameters appearing in
eqs. (\ref{c4:eqn:vertw}) and (\ref{c4:eqn:vertz}), rather than the
coefficients of the higher dimensional operators directly, in order to
be more general. The choice of operators is not unique, and different
choices will result in different translations between the two sets of
parameters (eqs. (\ref{paramsW}) and (\ref{paramsZ})). We discuss the
details of our analysis framework in the following section.

\section{Event Simulation and Selection}
\label{c6:sec:mc}
We consider $Vh$ production ($V=Z, W^\pm$), where the $V$ decays
leptonically, and the $h$ boson to a $b\bar{b}$ pair.
Further, we use jet substructure algorithm techniques to not just separate QCD backgrounds but use it to reconstruct the parton-parton CMS frame for $Vh$ production.
In this section we describe the tools and methods used for our
analysis, including the selection cuts utilised for both $Wh$ and $Zh$
production. We simulate all processes using {\tt
  MadGraph5}~\cite{Alwall:2011uj}, having implemented the effective
Lagrangian in {\tt
  FeynRules}~\cite{Christensen:2008py,Alloul:2013bka}. The output is
interfaced with {\tt Pythia6}~\cite{Sjostrand:2006za} for showering
and hadronization. We use the `Z2Star' tune for {\tt Pythia6},
including initial and final state radiation along with effects of
multiple interactions, and use the CTEQ6L1 parton distribution
functions~\cite{Pumplin:2002vw}. We use {\tt
  FastJet}~\cite{Cacciari:2011ma} to cluster jets.
Note that cross-checks were carried out at the parton level using analytic expressions.\\

Our selection
cuts for these signal processes are as follows.

\subsection{$Zh$ production}
For $Zh$ production we require:
\begin{enumerate}
\item A fat jet of radius $R=\sqrt{\Delta y^2 + \Delta \phi^2}=1.2$
  and transverse momentum $p_T>200\,{\rm GeV}$. After applying the
  mass drop and filtering procedure of~\cite{Butterworth:2008iy} on
  this fat jet, we require no more than three sub-jets with
  $p_T>20\,{\rm GeV}$, \mbox{$|\eta|<2.5$}, and radius
  $R_{sub}=\min(0.3,R_{bb})$, where $R_{bb}$ is the separation of the
  two hardest subjets, both of which must be $b$-tagged. In addition,
  we also require that the invariant mass of this jet system
  reconstructs the Higgs mass in the range \mbox{$110$ - $140\,$GeV}.
\item Exactly 2 leptons (transverse momentum $\pt>20\,{\rm GeV}$,
  pseudo-rapidity $|\eta| <2.5 $) of same flavour and opposite charge,
  with invariant mass within $10\,{\rm GeV}$ of the $Z$ mass
  $m_Z$. These should be isolated.
%  : the sum of all particle transverse
%  momenta in a cone of radius $R=0.3$ about each lepton should not
%  exceed 10\% of that of the lepton transverse momentum.
\item The reconstructed $Z$ has a $\pt >150\,{\rm GeV}$, with
  azimuthal angle satisfying $\Delta\phi(Z,h)> 1.2$.
\end{enumerate}
The first selection requirement listed above is used to reconstruct
the decaying Higgs. The requirement for a fat jet with large
transverse momentum means that we are looking at events with a highly
boosted Higgs. Note that by allowing for a third hard jet inside the
fat jet, the procedure allows for an extra jet other than the two
b-jets originating from the radiation of a gluon from the
b-quarks. The second requirement allows for reconstruction of the Z
boson, where the isolation criterion removes most of the $t\bar{t}$
and QCD background. The third requirement ensures that the Higgs and
the Z boson lie in almost the same plane of production as is expected
for the signal.
Note that the above selection criteria are applied after simulating a detector response to be described in section~\ref{subsec:Delphes}.
\\

After cuts, the only significant surviving background process is $Z$ +
jets.  Cross-sections at Leading Order (LO) after cuts are shown in
table~\ref{tab:cs1}. The $h\to b\bar{b}$ branching ratios were taken
from Ref.~\cite{Dittmaier:2011ti}. The cross-sections for backgrounds,
the SM, the pure CP--odd operator $c_V\neq0$ and the two BSM CP--even
operators $\bav\neq 0$ and $\bbv\neq 0$ are shown, with all other
couplings set to zero in each case. The values of the BSM couplings
are chosen so as to reproduce the SM total cross-section, without any
selection cuts. Note that in spite of this choice of couplings, the
cross-sections after cuts for the BSM cases are much higher than the
SM, implying a greater acceptance for the BSM cases. We will elaborate
on this point later.

\begin{table}
\centering \small
\begin{tabular}{|c|c|c|c|c|c|c|c|c|}\hline
Channel & $Vh_{SM}$ & V+jets & $\ttbar$ & Single top &
$\baw=0.2$ & $c_W=0.25$ & $\bbw=0.03$ \\ \hline
$W^{\pm}h$ & 0.355 & 0.28  & 0.13 & 0.06 & 1.45 & 2.14 &  7.11 \\ \hline                             
     &      &      &      &      &$\baz=0.23$    & $c_Z=0.30$    & $\bbz=0.08$    \\ \hline 
$Zh$ & 0.12 & 0.23 & 0    & 0    & 0.48 & 0.73 & 2.22 \\\hline

\end{tabular}
\caption{Cross-sections (fb) evaluated at leading order for the
  $14\,{\rm Te}$V LHC after applying all cuts. $V$+jets corresponds to
  the $Z$+jets background for the $Zh$ process and $W$+jets for the
  $Wh$ process. For the last three columns the SM contribution was set
  to zero and each of the values of $\bav,\bbv,c_V$ were set to
  reproduce the SM total cross-section before applying cuts.}
\label{tab:cs1}
\end{table}

\subsection{$Wh$ production}
For $Wh$ production we require the following:

\begin{enumerate}
\item The Higgs reconstructed as above.
\item Exactly one hard lepton ($\pt > 30\,{\rm GeV}$, $|\eta| <2.5 $),
  isolated as above.
\item Missing transverse momentum $\ptm > 30\,{\rm GeV}$.
\item The reconstructed $W$ has $\pt > 150\,{\rm GeV}$ and azimuthal
  angle satisfying $\Delta\phi(h,W) > 1.2$.
\item No additional jet activity with $\pt^{jet}>30\,{\rm GeV}$, and rapidity
  $|y|< 3$ (to suppress single and top pair production
  backgrounds).
\end{enumerate}
The difference between $Zh$ and $Wh$ is that in the latter case only
the transverse momentum of the W boson can be determined in the
detector. However, it is possible to some extent to reconstruct the
neutrino momentum as will be discussed in
what follows. The LO cross-section for the signal and major
backgrounds are detailed in~table~\ref{tab:cs1}. Once again, the
choice of couplings $(\baw,\bbw,c_W)$ is such that the total
cross-section (before any kinematic cuts) is identical to the SM total
cross-section. As in the case of $Zh$, we see that in this case also,
the cross-section after cuts is larger for the BSM couplings,
indicating a higher acceptance of the selection cuts to BSM physics.
In the following we describe certain detector effects that we have
considered in our analysis.

\subsection{Detector effects}
\label{subsec:Delphes}
While a full detector simulation is beyond the scope of this study, it
is still important to check whether the inefficiencies of a detector
do not dilute the effects that are observable with exact
reconstruction. 
To this end we use the {\tt Delphes 3} package~\cite{deFavereau:2013fsa} for a fast simulation 
of detector response. 
We set the parameters of the detector simulation tuned for the CMS detector, with some modifications:
\begin{itemize}
\item The lepton isolation radius R is set to a reduced value of 0.3, to allow for isolation of leptons in high transverse momentum events where the leptons will be collimated.
\item We modify the jet reconstruction algorithm for the detection of a boosted Higgs as described above
	and set the b-tagging efficiency to be $0.6$ while the mis-tagging efficiency for c-jets is $0.2$ and for all other jets is $0.001$.
\end{itemize}	
We use the {\tt Delphes} package since it has been shown to give good agreement with data~\cite{deFavereau:2013fsa}.
However, we do not carry out any validation with experiment for our choice of parameters as this is beyond the scope of the discussion presented here.

\subsection{Higher order effects}
At LO and at NLO in QCD, $Vh$ production occurs through
quark-initiated processes. The NLO (QCD) correction to the LO order
process is given entirely by corrections to the Drell-Yan
process~\cite{Han:1991ia,Baer:1992vx,Ohnemus:1992bd}. The NLO process
produces extra QCD radiation in the initial state thus affecting
observables such as the transverse momentum of the final state
particles. It should be noted that such effects are large only near
the threshold of the transverse momentum cuts due to collinear and/or
soft initial state radiation~\cite{Dittmaier:2012vm}. In fact the use
of asymmetric transverse momentum cuts on the V-boson ($p_T^V>150$ GeV)
and Higgs ($p_T^h>200$ GeV) transverse momenta means that most of this
effect will be concentrated in the region ($p_T^V<200$ GeV). 
However, the contribution to the cross-section from this region of phase space is small and hence we neglect this effect.\\

The K-factor (ratio to the LO order cross-section) for NLO (QCD) is
about $1.2$, see for example ref.~\cite{Dittmaier:2012vm}. In the special kinematic region of the boosted analysis
the K-factor for both $Zh$ and $Wh$ was found to be $\sim 1.5$ in
Ref.~\cite{Butterworth:2008iy}. For the $Zb\bar{b}$ background the
K-factor was found to be about the same while for $Wb\bar{b}$ the K
factor is higher and about $2.5$. The other main background,
$t\bar{t}$ production, was found to have a K-factor $~\sim 2$.  We
will use these values of the K-factor in our analysis of likelihoods in section~\ref{c6:sec:ll}.\\

Furthermore, we simulate the kinematics of an extra jet using the MLM
matching procedure~\cite{Hoche:2006ph} with one additional jet for
both signal and background. We have checked that our results do not
vary significantly with the addition of this extra jet. For $Wh$
production, since we veto events with additional hard jets to remove
backgrounds, the effect of extra radiation on the observables we
consider is negligible.

\subsection{\label{sec:w-reco}Reconstructing the neutrino momentum}

One must reconstruct the neutrino in $Wh$ production to determine our
angular observables.  We identify the neutrino transverse momentum
$\vec{p}_{T\, \nu}$ with the missing transverse momentum $\vec{\ptm}$.
As explained previously, the missing transverse momentum is
approximated by taking the negative vector sum of the transverse
momentum of all particles that can be detected ($> 0.5\,$GeV). In
order to evaluate the full four momentum of the neutrino, we demand
that the squared sum of the neutrino and lepton momenta be equal to
the squared $W$ boson mass $((p_{\nu}+p_{l_1})^2=M_{W}^2)$, and solve
the resulting quadratic equation.  Comparing with the ``true''
Monte-Carlo generated neutrino momentum, we find that choosing a given
solution out of the two possible ones, reconstructs the true neutrino
momentum $50\%$ of the time, with $\simeq 5\%$ giving imaginary
solutions. One may improve on this by comparing the boosts of the
Higgs $\beta_{z}^{h}$ and reconstructed $W$ $\beta_{z}^{W}$ in the $z$
direction. The solution with the minimum value for $|\beta_{z}^{W} -
\beta_{z}^{h}|$ gives the true neutrino momentum in $65\%$ of cases.
We thus present all our results using the latter algorithm.

\subsection{Sensitivity to anomalous couplings}
\label{c6:sec:sens}

%==================================
\begin{figure}
\begin{center}
\includegraphics[scale=0.27]{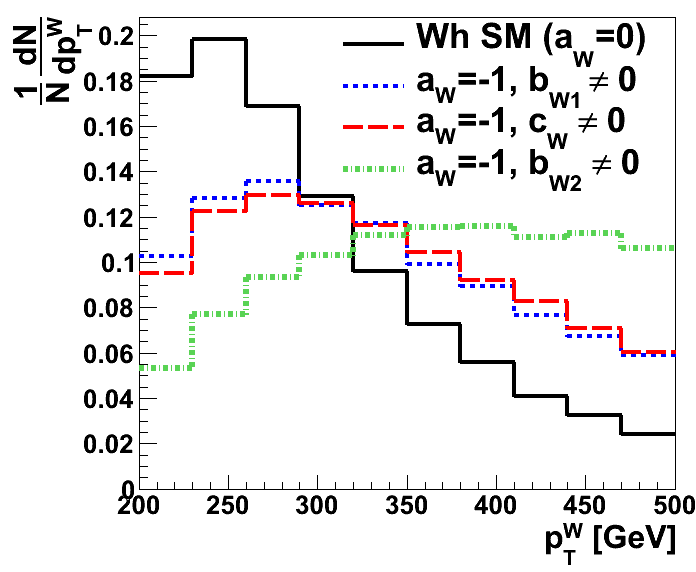} 
\includegraphics[scale=0.27]{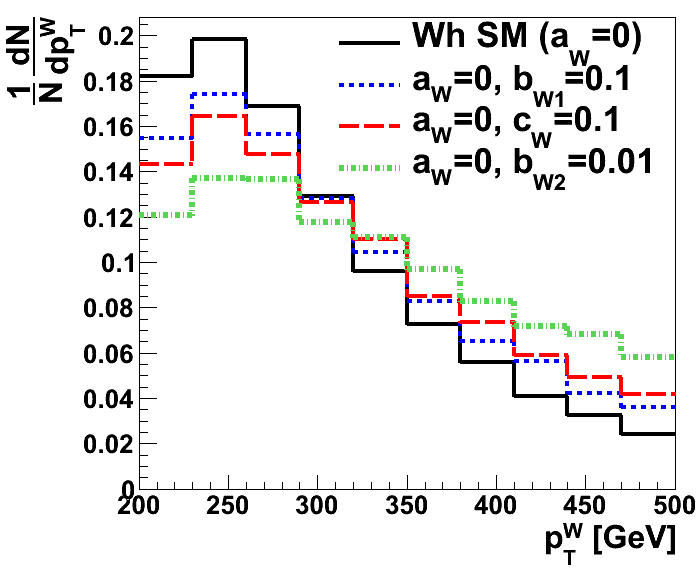}
%\includegraphics[height=3.8cm]{figs/vh_detaHV.eps}
%\scalebox{0.8}{\includegraphics{fig6_1.png}}
\end{center}
\caption[]{Plots showing the transverse momentum dependence of the W boson in $Wh$ production after applying the selection cuts listed in~\sref{c6:sec:mc}.
\textbf{Left}: Pure SM or BSM (all other couplings zero);
\textbf{Right}: Three cases of admixtures of the SM and BSM couplings (all other couplings set to zero). }
\label{c6:fig:pt}
\end{figure}
%==================================

It has been observed that the momentum dependence of the BSM couplings
of the $hVV$ vertex push the $\pt$ and invariant mass
$(\sqrt{\hat{s}_{Vh}})$ distributions to larger
values~\cite{Djouadi:2013yb,Ellis:2012xd,Englert:2012ct}, due
ultimately to the extra momentum factors present in the BSM
vertices. This is confirmed in the distributions shown in the plots in
\fref{c6:fig:pt}. The plot on the left shows the transverse momentum
distribution of the $W$ boson in $Wh$ production in the SM (black
solid line) compared each of the pure BSM couplings (with the SM
contribution set to zero), using the selection cuts described in the
previous section. The values of the couplings have been chosen so that
they reproduce the SM cross-section when no cuts are applied. We see
that the effect of all the BSM couplings is to push the $\pt^W$
distribution to larger values. The effect is even more pronounced for
the coupling $\bbw$ while the $\baw$ and $c_W$ couplings have a strong
but less pronounced effect on this distribution. In the right plot we
show the same distribution but this time for admixtures of the SM
coupling ($a_W=0$) with each of the BSM couplings. The effect of the
BSM couplings is still easily discernible, though less prominent than
compared to the plot on the left.  The larger $\pt$ distributions of the
$Vh$ system also lead to larger Higgs boosts and a reduced separation
$R_{ll}$ and $R_{bb}$ between the leptons (from the decay of the gauge
bosons) and $b$ jets (from the decay of the Higgs) respectively. As
mentioned in the previous section, this effect is further quantified
by the results of table~\ref{tab:cs1}, in which the cross-section for
pure BSM processes after cuts is significantly higher than the SM
result, after imposing that the cross-sections agree before cuts.\\

\begin{figure}
\centering
\includegraphics[width=0.7\textwidth]{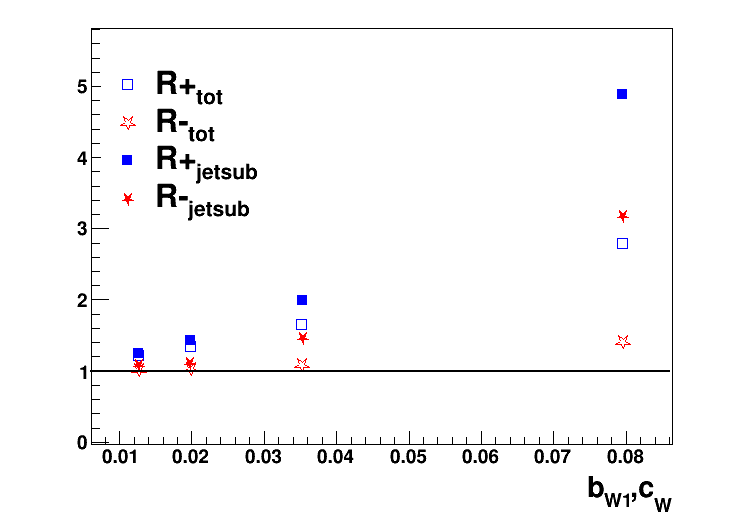}
%\scalebox{0.8}{\includegraphics{fig6_2.png}}
\caption{The ratio of the cross-sections $R-$ (mixture of SM and CP
  odd) (red stars) and $R+$ (mixture of SM and the BSM CP even term)
  (blue boxes) both before (hollow markers) and after (bold markers)
  applying selection cuts for the $Wh$ channel for $14$~TeV LHC. The $x$-axis corresponds to the strength of each of the couplings
  $\baw$ or $c_W$.
}
\label{c6:fig:ratio}
\end{figure}

In~\fref{c6:fig:ratio}, we consider the SM coupling $a_W=0$
supplemented by either the $c_W$ (CP--odd) coupling or the $\baw$
coupling applied to the $Wh$ channel.  We show the ratio of the SM+BSM
and SM cross-sections both for the total cross-section
\mbox{$(R_{tot}^{\pm}=\sigma_{tot}^{SM+BSM\pm}/\sigma_{tot}^{SM})$}
and the cross-section after applying selection cuts
\mbox{$(R_{jetsub}{\pm}=\sigma_{jetsub}^{SM+BSM\pm}/\sigma_{jetsub}^{SM})$}.
Here $R^{+}$ and $R^{-}$ correspond to the case ($a_W=0,\baw\neq 0$)
and ($a_W=0,c_W\neq 0$) respectively.  As is expected, both ratios
decrease with the strength of the BSM couplings and approach unity as
the BSM couplings tend to zero.  $R_{tot}^{+}$ shows a faster rise
with coupling strength than $R_{tot}^{-}$. This is because the
interference term in the matrix element squared for the CP--odd
coupling does not contribute to the cross-section\footnote{In
  practice, if the squared term is larger than the interference term,
  this is an indication that the effective theory framework is
  breaking down. Here, however, we are merely quantifying the effect
  by which selection cuts enhance BSM effects, by fixing the BSM
  cross-section to be artificially high (equivalent to an unphysically
  low cut-off scale).}.  Importantly, $R_{jetsub}$ (for both
couplings) increases at a faster rate than $R_{tot}$ with increasing
values of the corresponding couplings.  Similar results also hold for
the second CP--even anomalous coupling $\bbw$.  These ratios are
therefore quite sensitive to the presence of anomalous
couplings. Whilst not directly experimentally measurable, they can be
determined by comparing an experimental measurement of the $Vh$ signal
with a precise theoretical prediction for the SM only contribution. If
this lies away from unity, this constitutes a strong indication of BSM
physics. Another feature that should be noted is that the ratio of the
ratios ($R^{\pm}_{jetsub}/R^{\pm}_{tot}$) increases at a faster rate
for the CP--odd coupling than it does for the CP--even coupling. This
is in agreement with the results of \tref{tab:cs1}, where it was
observed that the acceptance to the selection cuts of the
pseudo-scalar state was higher than a scalar with anomalous coupling
$\bbw$.

\section{Angular observables}
\label{c6:sec:ang}

In this section, we consider differential observables that can
distinguish between the different BSM vertices occurring in the $hVV$
interaction. One such observable, the transverse mass of the $Vh$
system, has been used at the Tevatron to probe the $hWW$
vertex~\cite{Johnson:2013zza}, however, it has been shown to be
ineffective at the LHC~\cite{Ellis:2012xd}. Furthermore, the CP-odd
coupling contributes to this observable only through quadratic terms
in the matrix element squared and therefore is not the most sensitive
observable~\footnote{Care must be taken if such quadratic terms become
  important, as this signals a potential breakdown of the effective
  theory description.}. Angular observables, as we will show, can be
\textit{linearly} sensitive to the anomalous couplings. This is useful in that
one may construct asymmetry parameters that are manifestly zero for
the SM, such that any non-zero measurement constitutes discovery of
new physics. Note that in the context of effective theory analysis, constructing
observables that are linear in the anomalous couplings is of paramount importance.\\

The tensor structure of the BSM vertices will be reflected in the
angular distribution of the decay products of the gauge
boson\footnote{Note, the Higgs has spin zero; angular distributions of
  its decay products are uncorrelated from production and hence do not carry any information about the $hVV$ vertex in $Vh$ production. Angular correlations of the decay products of the Higgs would reflect the properties of the decay vertex and not the production vertex; as in $\dhvv$ decays.}. To this end, we construct various angular observables that
could discriminate between the different vertex structures.  The
momenta of the $V$ and Higgs bosons are reconstructed from the leptons
and jets as follows:
\begin{equation}
p_V=p_{l_1}+p_{l_2}, \quad p_h=p_{b_1}+p_{b_2} + p_{j},
\label{ZHmoms}
\end{equation}
where  $\{p_{b_i}\}$ are the momenta of the $b$ jets, $p_{j}$ is
the momentum of the third jet if it is reconstructed and $p_{l_1}$ and
$p_{l_2}$ are the momenta of the lepton and the anti-lepton
respectively (for $Wh$, $p_{l_1}$ corresponds to the lepton momentum
and $p_{l_2}$ to the neutrino). With these momenta, we may define

\begin{equation}
\cos\theta^*=
\frac{\vec{p}_{l_1}^{\,(V)}\cdot\vec{p}_V}{|\vec{p}_{l_1}^{\,(V)}|\,|\vec{p}_V|},
\label{c6:eqn:tstar}
\end{equation}
Here $\vec{p}_{X}^{\,(Y)}$ corresponds to the three momentum of the
particle $X$ in the rest frame of the particle $Y$. If $Y$ is not
specified then the momentum is defined in the lab frame.  The
parameter $\cos\theta^*$ corresponds to the angle between the
direction of the decaying lepton in the rest frame of the V boson with
the direction of flight of the V boson in the lab frame.  This angle,
first defined in \cite{Miller:2001bi}, encodes the $W$ boson
polarization.\\

\begin{figure}
\centering
\includegraphics[scale=0.4]{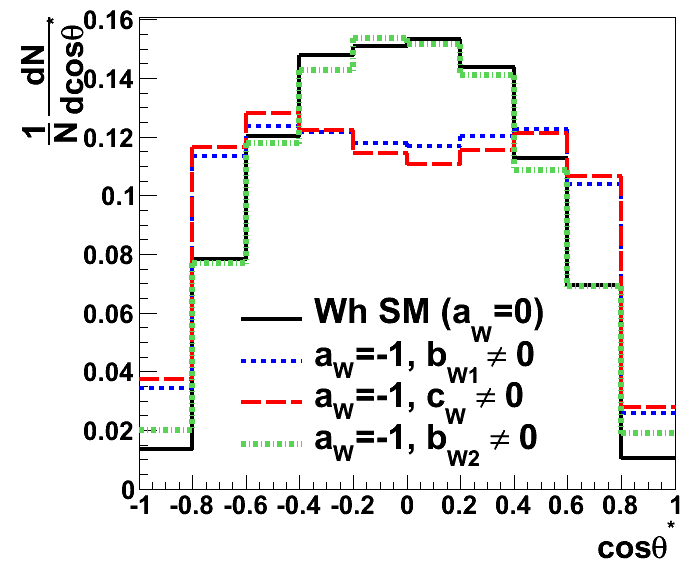}
%\scalebox{0.7}{\includegraphics{fig6_3.png}}
\caption{Plot of the distribution of the angle $\cos\theta^{*}$ for
  $W^{+}h$ production, for pure SM and BSM operators. The values of
  the couplings are chosen as in \tref{tab:cs1}, so as to reproduce
  the SM cross-section (before cuts).
\label{c6:fig:tstarw}
}
\end{figure}
The SM and BSM couplings lead to mostly longitudinal and transverse
$W$ bosons respectively. That $\cos\theta^*$ then effectively
distinguishes SM and BSM effects can be seen in \fref{c6:fig:tstarw}.
We see that the SM distribution (black solid line) peaks at
$\cos\theta^{*}=0$ and vanishes at $\cos\theta^{*}=\pm 1$. The
distribution for the BSM coupling $\bbw$ (green dot-dashed line)
closely follows the SM distribution. This is expected since the tensor
structure of the vertex in both these cases is the same. In contrast,
the couplings $\baw$ and $c_W$ produce distributions that have minima
at $\cos\theta^{*}=0$. They too appear to vanish at
$\cos\theta^{*}=\pm 1$, however this is the effect of the selection
cuts\footnote{All the selection cuts deplete this region of phase space, with strongest effects coming from the transverse momentum and rapidity cuts of the lepton.}. Without applying any selection cuts the distribution for these
two cases peak at $\cos\theta^{*}=\pm 1$.\\

The behaviour of this distribution for each of the couplings can be
understood as follows.  For a transversely polarized W boson, the
decay lepton spins align themselves perpendicular to the direction
of motion of the W boson and gives rise to a distribution of the form
$(1\pm \cos\theta^{*})^2$, while in the case of a longitudinally
polarized W boson, the spins of the decay leptons align themselves
along the direction of the W boson and give rise to a distribution
of the form $\sin^2\theta^{*}$. The two BSM couplings $\baw$ and $c_W$
produce more transversely polarized W boson states while the SM
coupling and the coupling $\bbw$ produce more longitudinally polarized
W bosons. Using this distribution, it is therefore possible to
differentiate between vertex structures with couplings $\baw$ or $c_W$
from the vertex structure with couplings $a_W$ or $\bbw$, but not
between $\baw$ and $c_W$ or between $a_W$ and $\bbw$.\\

\begin{figure}
\centering
\includegraphics[scale=0.4]{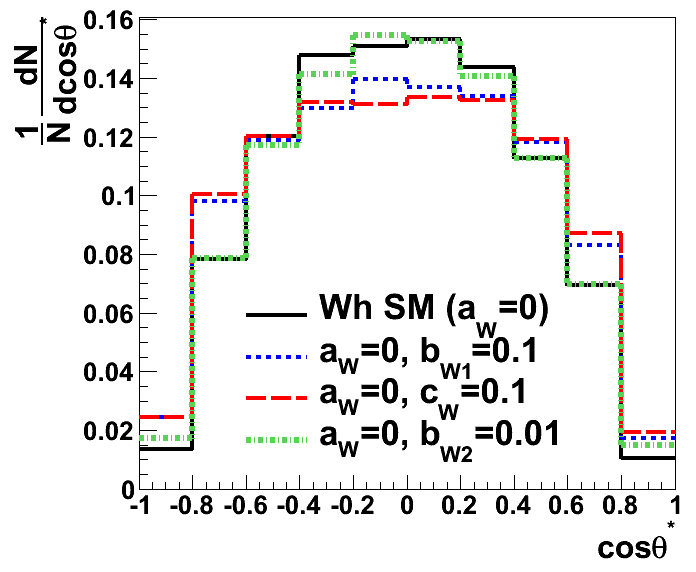}
%\scalebox{0.7}{\includegraphics{fig6_4.png}}
\caption{Plot of the distribution of the angle $\cos\theta^{*}$ for
  $W^{+}h$ production, for admixtures of SM and BSM couplings.
\label{c6:fig:inf_tstarw}
}
\end{figure}

In \fref{c6:fig:inf_tstarw} we present plots of the same observable
$\cos\theta^{*}$ for admixtures of the SM coupling with each of the
BSM couplings.  We present three cases corresponding to
($a_W=0,\baw=0.1$), ($a_W=0,c_W=0.1$) and ($a_W=0,\bbw=0.01$). We see
that differences, though reduced, are still discernible in this
distribution. Similar results also hold for $Zh$ production.\\

\begin{figure}
\centering
%\vspace{-1cm}
\includegraphics[scale=0.8]{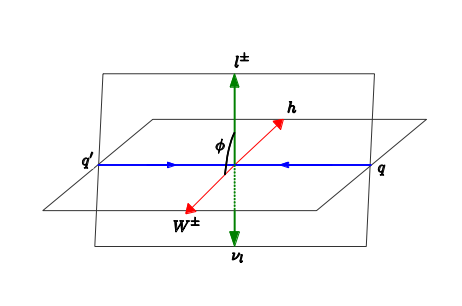}
%\scalebox{0.7}{\includegraphics{fig6_5.png}}
\caption{The angle ($\phi$) between the plane of production and the
  lepton from the decay of the W boson.
\label{c6:fig:schem}
}
\end{figure}

To fully distinguish the CP even ($\baw$) and odd ($c_W$) BSM
contributions, one must construct CP-odd observables, which is
difficult in principle for a proton--proton
collider~\cite{Han:2009ra}. For $Zh$ production,
Ref.~\cite{Christensen:2010pf} considered two such observables,
although these are sensitive to radiation and hadronization
corrections; Ref.~\cite{Englert:2012xt} defined observables which are
insensitive to the CP nature of BSM contributions.
Ref.~\cite{Desai:2011yj} examined CP--odd asymmetries in $Wh$
production with the decay $h \to W^{(*)} W^*$, though the effect of
the BSM CP even term was not considered. The hint for possible CP--odd
observables comes from looking at the matrix element squared for the
process $q(k_1) \bar{q}^{\prime}(k_2)\to W^{+}(p_W) h(p_h) \to
l^{+}(p_1) \bar{\nu}_l(p_2) h(p_h)$ shown in~\apref{c6:app}. The
interference term between the CP--odd coupling and the SM coupling
($a_Wc_W$), is proportional to
$\epsilon_{\mu\nu\rho\sigma}k^{\mu}(p_h-k_1)^{\nu} p_{W}^{\rho}
p_1^{\sigma}$, where $k=k_1+k_2$ and $\epsilon_{\mu\nu\rho\sigma}$ is
the anti-symmetric Levi-Civita tensor. Such a term depends on the
angle between the plane of production of the $Wh$ and the direction of
flight of the lepton. This is depicted in \fref{c6:fig:schem}.\\

We now construct the following angles based on the observation made above.

\begin{eqnarray}
\cos\delta^+&=&
\frac{\vec{p}_{l_1}^{\,(V)}\cdot\left(\vec{p}_h\times \vec{p}_V\right)}
{|\vec{p}_{l_1}^{\,(V)}|\,|\vec{p}_h\times \vec{p}_V|}, \nonumber \qquad
\cos\delta^- = 
\frac{(\vec{p}_{l_1}^{\,(h-)}\times \vec{p}_{l_2}^{\,(h-)})\cdot\vec{p}_V}
{|(\vec{p}_{l_1}^{\,(h-)}\times \vec{p}_{l_2}^{\,(h-)})||\vec{p}_V|},\\ \\ 
\Delta\phi^{lV}&=&\Delta\phi(\vec{p}_{l_1}^{\ (V)},\ \vec{p}_V).
\label{c6:eqn:angles}
\end{eqnarray}

We use the same notation for the momenta as described below
\eref{c6:eqn:tstar}.  For an $e^{+}e^{-}$ collider where the direction
(as well as the energy) of the lepton and anti-lepton are well known,
it is sufficient to define the normal to the plane of production with
the cross-product between any one of the leptons and the direction of
flight of the Higgs or gauge boson. Note that the choice of vectors in
$e^{+}e^{-}$ collisions completely fixes whether the normal points
`below' or `above' the plane of production. At the LHC, the
information about the direction of the quark or anti-quark is not
known and hence it is difficult to fix the direction of the normal to
the plane of production. However, it is known that the valence quarks
are likely to carry a larger fraction of the proton momentum. The
direction of the normal to the plane of production can then be fixed
by the momenta of the V and Higgs bosons.
\begin{figure}
\centering
\includegraphics[scale=0.29]{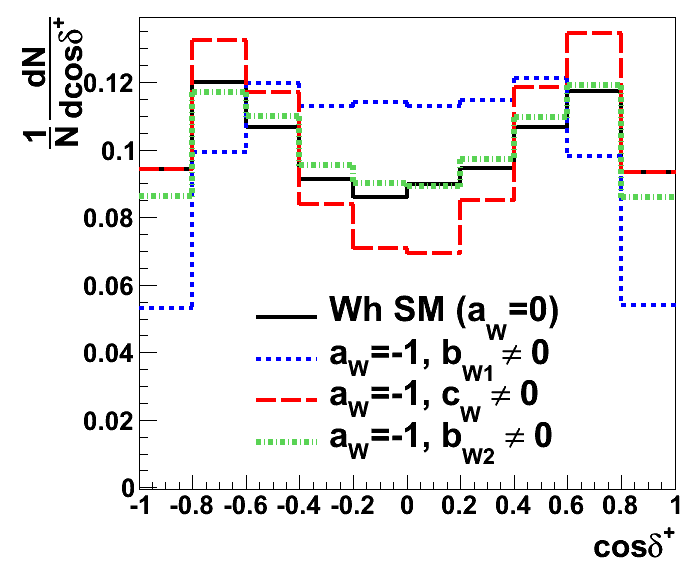}
\includegraphics[scale=0.29]{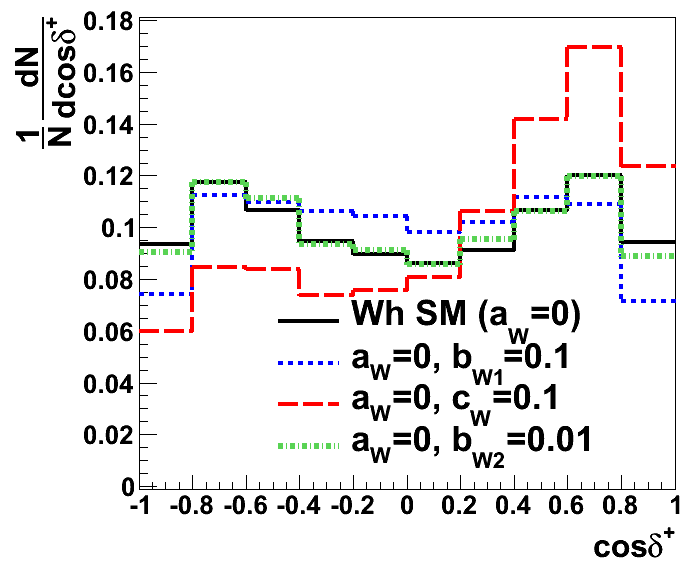}
%\scalebox{0.7}{\includegraphics{fig6_6.png}}
\caption{Plots showing the distribution of the angle $\cos\delta^{+}$
  defined in \eref{c6:eqn:angles}. \textbf{Left}: Pure SM and BSM
  couplings, chosen as in \tref{tab:cs1}, so as to reproduce the SM
  cross-section (before cuts).  \textbf{Right}: Various admixtures of
  SM and BSM operators, with all other couplings set to zero.
\label{c6:fig:delpw}
 }
\end{figure}
We use this fact to construct the first angle in \eref{c6:eqn:angles},
$\cos\delta^{+}$, which corresponds to the angle between the direction
of flight of the lepton (with the momentum evaluated in the rest frame
of the V boson) and the plane formed by the V boson and the Higgs. In
an $e^{+}e^{-}$ collider, where the centre-of-mass and lab frames
coincide, the gauge and Higgs bosons will be produced back to back.
However, at the LHC one can take advantage of the asymmetric collision
energies of the partons which results in the difference between the
centre-of-mass and lab frame. For asymmetric collisions, the plane
defined by the cross-product between the V boson and Higgs directions,
coincides with the plane of production. In \fref{c6:fig:delpw} we show
the distribution of this observable for SM as well as for the
anomalous couplings. In the plot on the left, the SM prediction is
compared to the prediction for each of the anomalous couplings : $\baw
\neq 0$ (blue dotted line), $\bbw$ (green dot-dashed) and $c_W$ (red
dashed) with all other couplings set to zero in each case. All
distributions show a dip at $\cos\delta^{+}=0$. This is created by the
transverse momentum cut on the leptons, since low $\pt$ leptons will
always be perpendicular to the normal to the plane of production.\\

We see from this distribution that for $\baw\neq 0$ leptons are
produced mostly in the plane of production, while for $c_W\neq 0$ the
leptons tend to be produced mostly perpendicular to the plane of
production. For two cases, the SM and for $\bbw\neq 0$, the
distribution is flat (without cuts) and has a slight dip at
$\cos\delta^{+}=0$ due to the $\pt$ cuts, as explained above. This
observable clearly discriminates between the $\baw\neq 0$ and $c_W
\neq 0$ cases, a feature that was absent in the distributions of
observables discussed earlier.\\

More interesting effects can be seen in this observable when we
consider admixtures of each of the anomalous couplings with the SM. In
the right plot of \fref{c6:fig:delpw} the distribution of three cases
are compared with the SM expectation: $(a_W=0,\baw= 0.1)$ (blue dotted
line), ($a_W=0,\bbw=0.01$) (green dot-dashed) and $(a_W=0,c_W=0.1)$
(red dashed), with all other couplings set to zero in each case. As
expected, the distribution for the case ($a_W=0,\bbw=0.01$) follows
the SM distribution closely. The CP--even $(a_W=0,\baw= 0.1)$ case is
similar to the pure anomalous coupling case ($a_W=-1,\baw\neq 0$ ). We
have checked that the interference term alone for the case
$(a_W=0,\baw= 0.1)$ produces a similar distribution to the case
($a_W=-1,\baw\neq 0$) and is therefore linearly sensitive to $\baw$. For the
CP violating case $(a_W=0,c_W=0.1)$ the distribution is skewed towards
positive values of $\cos\delta^{+}$. This is due to the presence of
the Levi-Civita tensor in the interference term of the matrix element
squared as described above. Note that the distribution will peak for
negative values of $\cos\delta^{+}$ if the sign of the coupling $c_W$
were changed.  This observable is therefore, linearly sensitive
to $c_W$ and hence to its sign. We will use this fact to construct
asymmetries in the next section.\\

\begin{figure}
\centering
\includegraphics[scale=0.29]{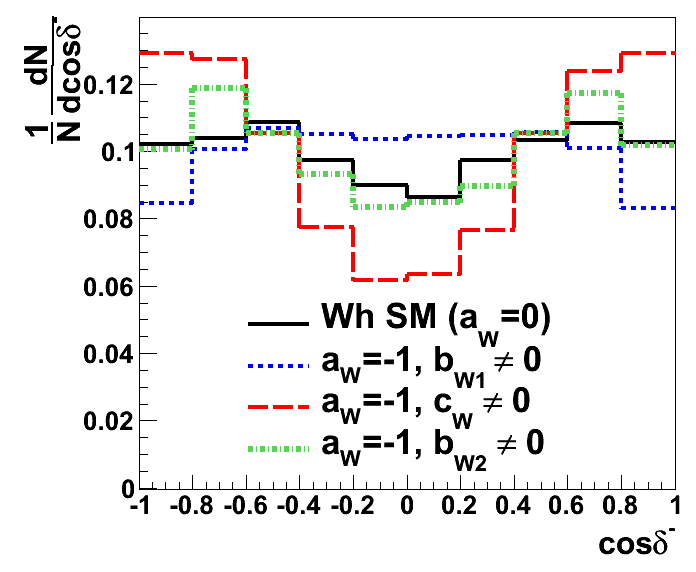}
\includegraphics[scale=0.29]{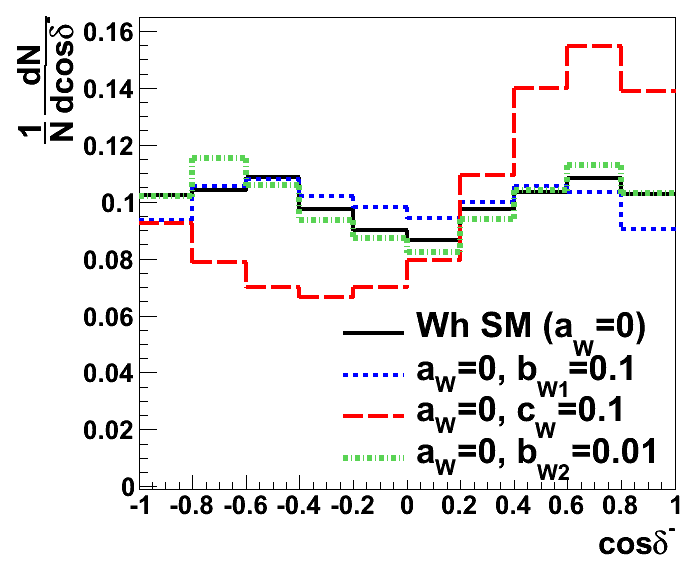}
%\scalebox{0.7}{\includegraphics{fig6_7.png}}
\caption{Plots showing the distribution of the angle $\cos\delta^{-}$
  defined in \eref{c6:eqn:angles}.  \textbf{Left}: Pure SM and BSM
  couplings, chosen as in \tref{tab:cs1}, so as to reproduce the SM
  cross-section (before cuts).  \textbf{Right}: Various admixtures of
  SM and BSM couplings, with all other couplings set to zero.
\label{c6:fig:delmw}
}
\end{figure}
The second observable we consider is slightly more complicated in
construction. The momenta of the two leptons from the decay of the
gauge boson are evaluated in the frame in which the Higgs would be at
rest, were its three momentum reversed\footnote{If the four momentum
  of the Higgs is written as $(E_h,\vec{p}_h)$, then this is the frame
  defined by the boost such that $(E_h,-\vec{p}_h) \to (m_h,0)$}. Then
$\cos\delta^{-}$ is the angle between the plane formed by the two
leptons in this frame and the V boson in the lab frame. This angle is
related to the angle $\phi$ depicted in \fref{c6:fig:schem}. The
distribution of this observable for SM is compared with three other
cases in the left plot of \fref{c6:fig:delmw} : $\baw \neq 0$ (blue
dotted line), $\bbw\neq 0$ (green dot-dashed) and $c_W\neq 0$ (red
dashed) with all other couplings set to zero in each case. The right
plot of \fref{c6:fig:delmw} compares the distribution of the SM
expectation with admixtures of the SM couplings and the BSM coupling
as given in the figure (and other couplings set to zero). The
behaviour of this angle is very similar to that of
$\cos\delta^{+}$. There are two noticeable differences. Firstly in the
case $c_W\neq 0$ the distribution of $\cos\delta^{-}$ appears to show
a more heightened difference from SM as compared to the distribution
for $\cos\delta^{+}$. For the case when $\baw\neq 0$ the opposite is
true and $\cos\delta^{+}$ appears to show a greater difference from
the SM distribution. This is also true when we set $(a_W=0,\baw=0.1)$. For
the CP-violating case, the same skewed behaviour of the distribution
that was observed for $\cos\delta^{+}$, reappears here. As usual the
distribution for the coupling $\bbw$ in both the pure and mixed cases
of the left and right plots of \fref{c6:fig:delmw}, follow closely the
SM expectation.\\

\begin{figure}
\centering
\includegraphics[scale=0.29]{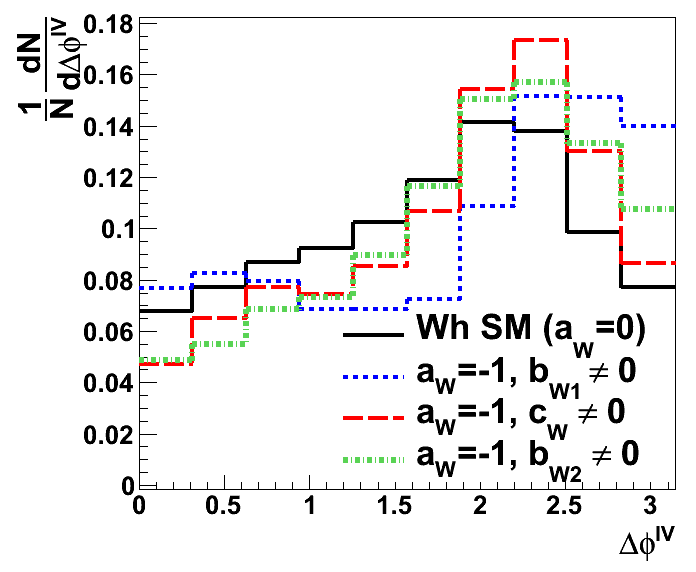}
\includegraphics[scale=0.29]{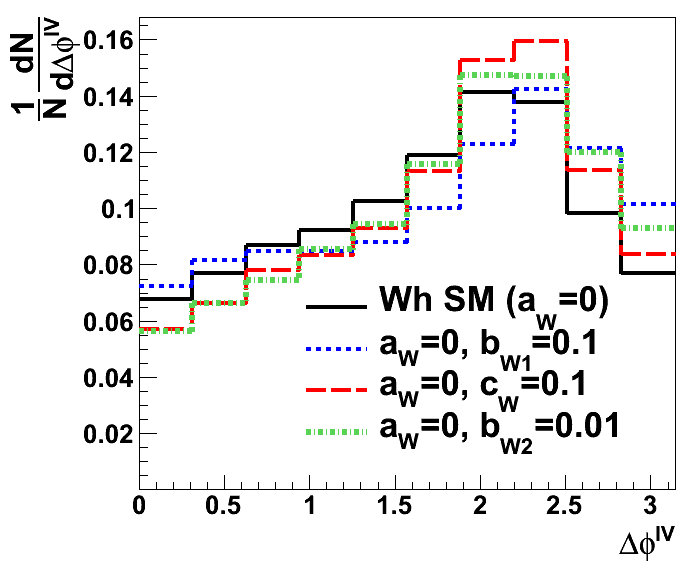}
%\scalebox{0.7}{\includegraphics{fig6_8.png}}
\caption{Plots showing the distribution of the angle $\Delta\phi^{lV}$
  defined in \eref{c6:eqn:angles}. \textbf{Left}: Pure SM and BSM
  couplings, chosen as in \tref{tab:cs1}, so as to reproduce the SM
  cross-section (before cuts).  \textbf{Right}: Various admixtures of
  SM and BSM couplings, with all other couplings set to zero.
\label{c6:fig:dphiw}
}
\end{figure}
The last observable we consider ($\Delta\phi^{lV}$) is the azimuthal
angle difference of the lepton momenta (evaluated in the rest frame of
the V boson) and the V boson momentum. The distribution for this
observable is shown in \fref{c6:fig:dphiw}. The left plot compares the
SM expectation with three cases of the pure BSM couplings. The right
plot compares the SM prediction with admixtures of SM with each of the
BSM couplings. For all the cases we consider, there is a significant
difference from the SM distribution of this observable. The most
striking difference, however, is for the pure CP--even case ($\baw\neq
0$) which displays a minimum in this distribution at $\Delta\phi^{lV}$
unlike the other cases. Differences between the distributions remain,
although reduced, when considering admixtures of the SM coupling
($a_W=0$) with each of the BSM couplings.\\

\begin{figure}
\centering
\includegraphics[scale=0.29]{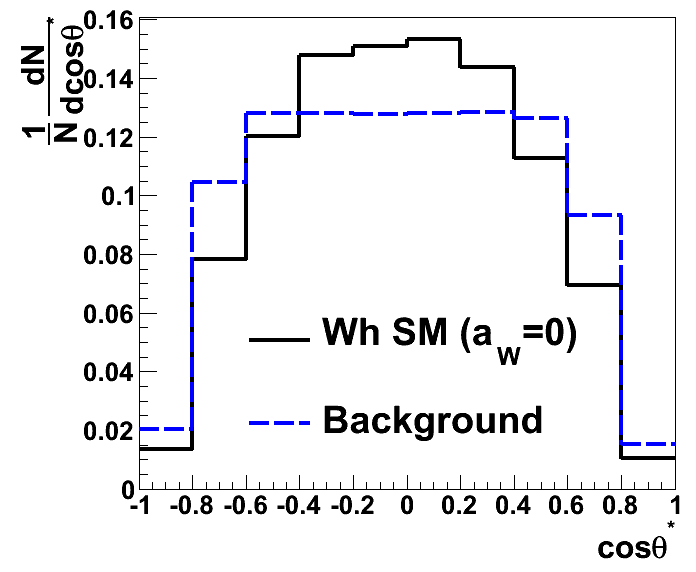}
\includegraphics[scale=0.29]{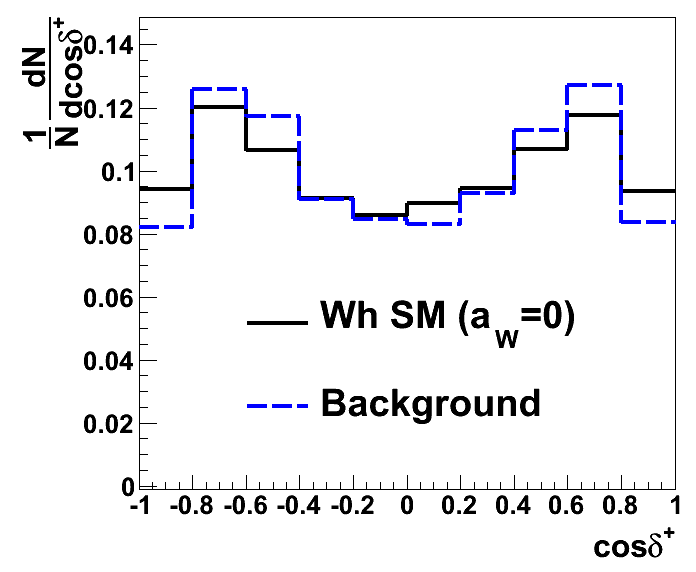}\\
\includegraphics[scale=0.29]{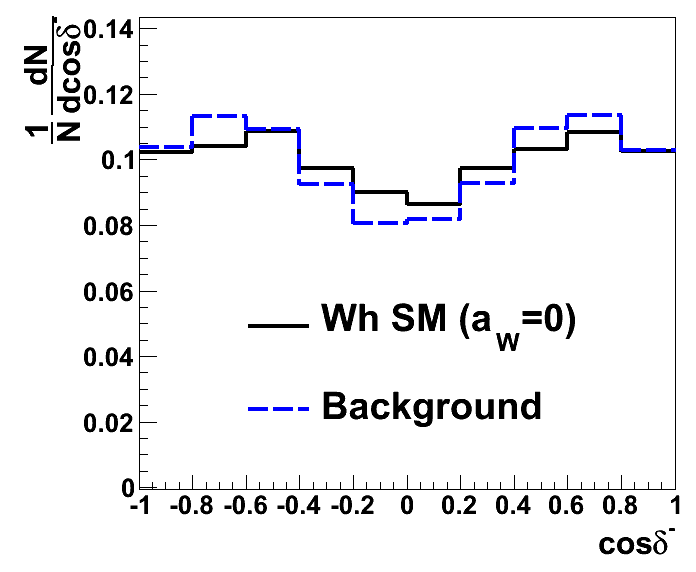}
\includegraphics[scale=0.29]{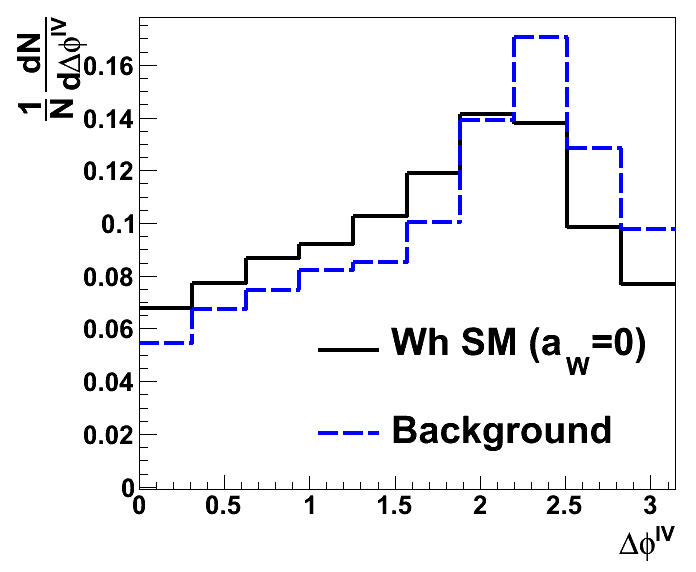}
%\scalebox{0.7}{\includegraphics{fig6_9.png}}
\caption{Normalized distribution of the observables defined in
  \eref{c6:eqn:tstar} and \eref{c6:eqn:angles} for the backgrounds
  (blue dashed) to $Wh$ production listed in \tref{tab:cs1} and SM
  predictions (black solid lines).
\label{c6:fig:bkg}
}
\end{figure}

\begin{figure}
\centering
\includegraphics[scale=0.29]{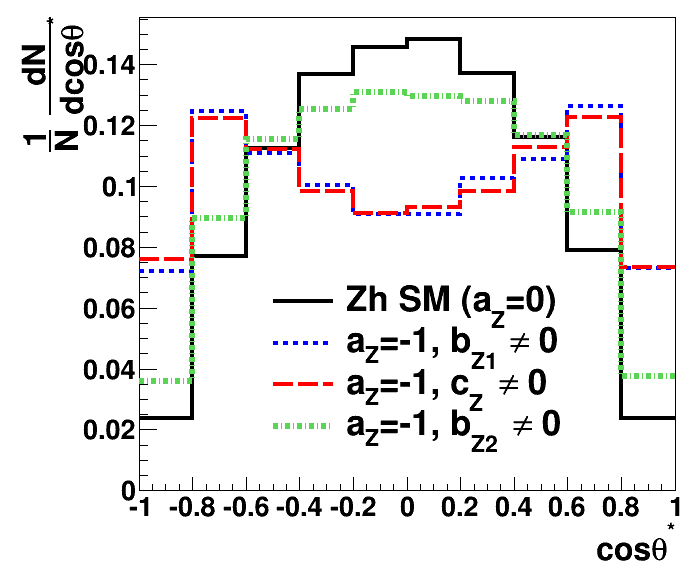}
\includegraphics[scale=0.29]{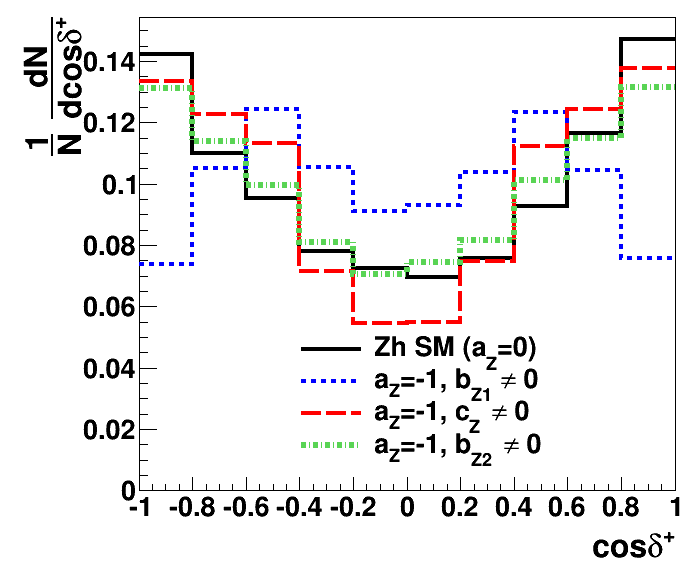}\\
\includegraphics[scale=0.29]{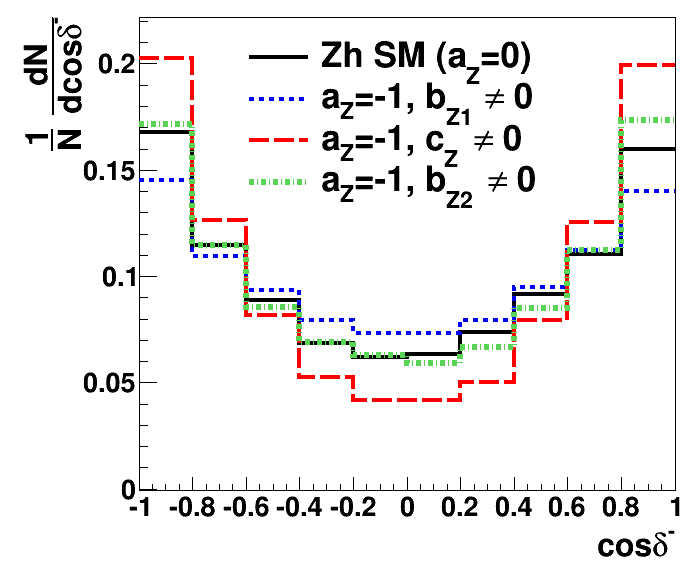}
\includegraphics[scale=0.29]{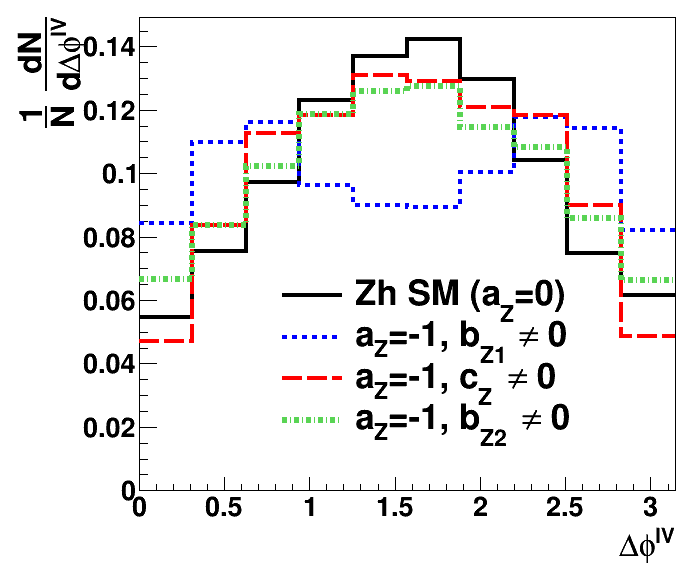}
%\scalebox{0.7}{\includegraphics{fig6_10.png}}
\caption{
Normalized distribution of the observables defined in
\eref{c6:eqn:tstar} and \eref{c6:eqn:angles} for $Zh$ production. 
The SM distribution (black solid line) is compared with the predictions of the three different BSM couplings: $(a_Z=-1,\baz \neq 0)$ (blue dotted line) with all other couplings set to zero, $(a_Z=-1,\bbz\neq 0)$ (green dot-dashed) with all other couplings set to zero and $(a_Z=-1,c_Z \neq 0)$ (red dashed) with all other couplings set to zero.
The values of the couplings are chosen as in \tref{tab:cs1}, so as to reproduce the SM cross-section (before cuts).
\label{c6:fig:an-z}
}
\end{figure}

We also show in \fref{c6:fig:bkg} distributions of the observables
described above for the backgrounds to $Wh$ production listed in
\tref{tab:cs1}. The distribution of the various angles follow the SM
distribution except for the angles $\Delta\phi^{lV}$ and
$\cos\theta^{*}$.
For completeness, we also show the distributions of the various angles
defined in \eref{c6:eqn:tstar} and \eref{c6:eqn:angles} for $Zh$
production. In \fref{c6:fig:an-z}, the SM distribution (black solid
line) is compared with the predictions of the three different BSM
couplings. The values of the couplings are chosen as in \tref{tab:cs1},
so as to reproduce the SM total cross-section (before cuts). The
distributions display a similar behaviour as compared to the analogous
distributions in $Wh$ production.\\

\begin{figure}
\centering
\includegraphics[scale=0.29]{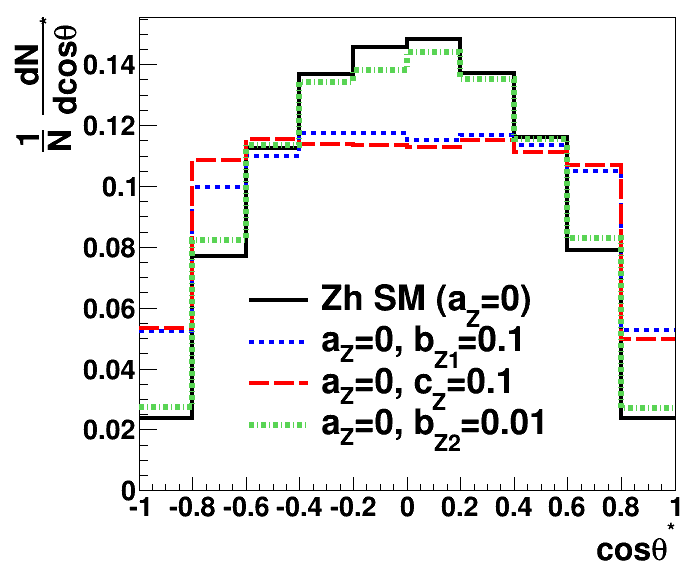}
\includegraphics[scale=0.29]{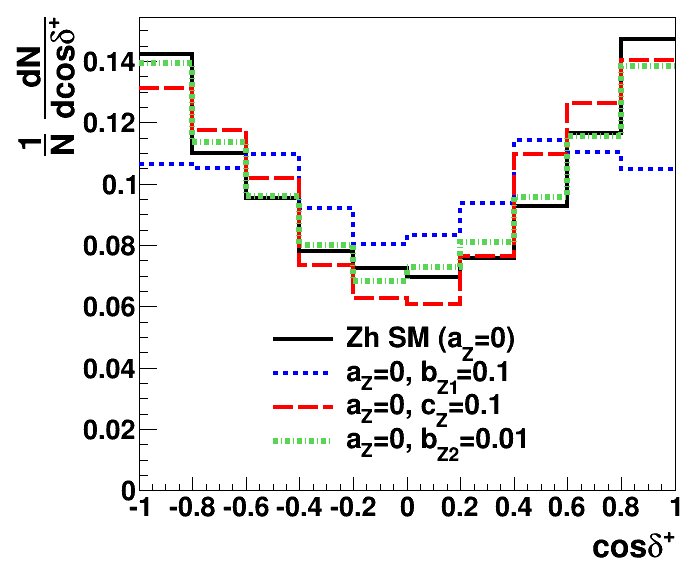}\\
\includegraphics[scale=0.29]{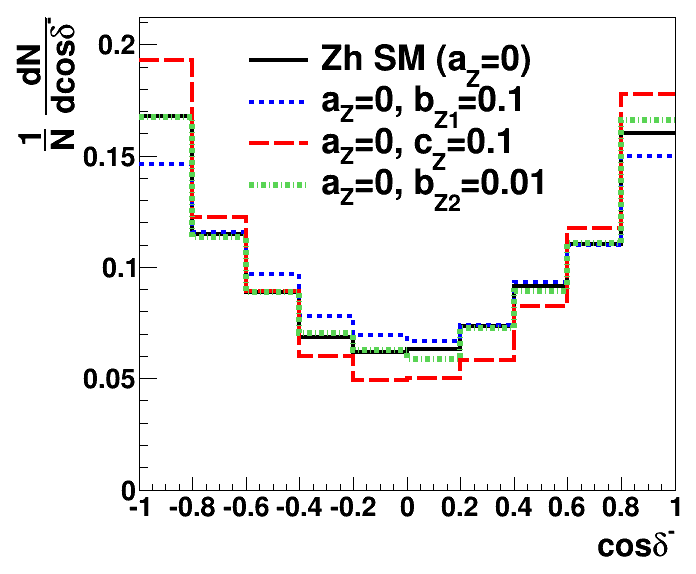}
\includegraphics[scale=0.29]{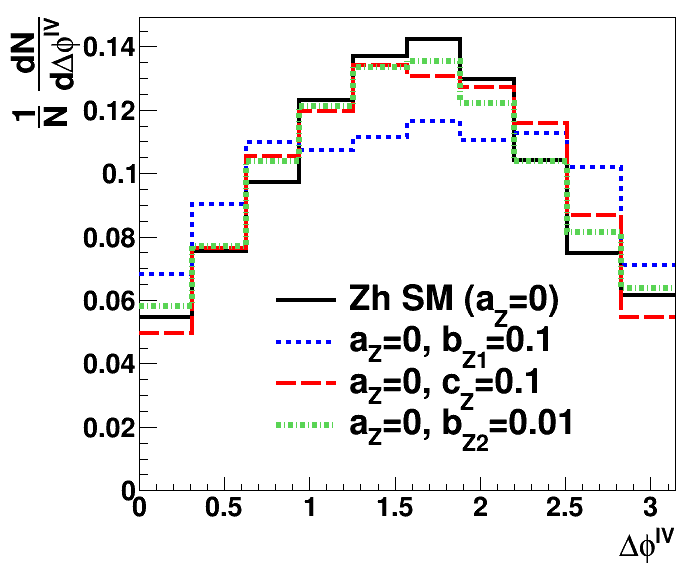}
%\scalebox{0.7}{\includegraphics{fig6_11.png}}
\caption{Normalized distribution of the observables defined in
  \eref{c6:eqn:tstar} and \eref{c6:eqn:angles} for the $Zh$
  production. Various admixtures of SM and BSM operators are shown,
  with all other couplings set to zero.
\label{c6:fig:inf_an-z}
}
\end{figure}

In \fref{c6:fig:inf_an-z}, the SM distribution (black solid line) is
compared with three cases which involve admixtures of the SM and BSM
couplings The asymmetries in the distributions of $\cos\delta^{+}$ and
$\cos\delta^{-}$ that one observes in $Wh$ production for the
CP-violating case $(a_Z=1,c_Z = 0.1)$, although present, are far less
prominent in $Zh$ production.  The reason for this difference can be
ascertained by looking at the CP violating term in the matrix element
squared. For $Wh$ production, as described earlier, this term was
simply proportional to a Levi-Civita tensor of the form
$\epsilon_{\mu\nu\rho\sigma}k^{\mu}(p_h-k_1)^{\nu} p_{W}^{\rho}
p_1^{\sigma}$.  The CP violating term in the matrix element squared
for $Zh$ production has several instances of the Levi-Civita tensor
that come with opposite signs. These do not cancel out (as they do in
$Wh$ production) since they are multiplied by axial and vector
couplings (which are of different strengths). As a result, the
distributions of $\cos\delta^{+}$ and $\cos\delta^{-}$ receive
contributions from Levi-Civita tensors of opposite sign and hence
display a reduced skewness in distribution in comparison to the
analogous distribution in $Wh$ production.\\

We now have a set of observables that can discriminate not only the SM
coupling from BSM couplings but also \textit{between} the various BSM
couplings, as evidenced by the distributions presented in this
section. In order to fully assess the discriminating power of these
observables and to estimate the typical luminosities that one would
require at a $14$~TeV LHC to rule out the various anomalous couplings,
we perform a multi-variable likelihood analysis in the next section.

\section{Multi-Variable Likelihood analysis}
\label{c6:sec:ll}
In the previous section we described the various observables that one
could use in order to probe anomalous couplings in $Vh$ production. We
found that the transverse momentum of the V boson (or the Higgs), the
angle $\cos\theta^{*}$ and any one of the correlated observables
defined in \eref{c6:eqn:angles} can be used for this purpose. It is
well known that the maximized log likelihood ratio provides the
strongest test statistic according to the Neyman-Pearson
lemma. Therefore in order to assess the sensitivity of these
observables to probe anomalous couplings at the LHC, we perform a
three dimensional extended binned-likelihood analysis. The procedure
we follow is outlined below.\\

We set the SM expectation ($a_W=0$) plus backgrounds as our null hypothesis. The
alternate hypotheses are chosen to be the various cases which involve
any one of the BSM couplings along with backgrounds.  We define our likelihood as functions
of a set of three observables. These are $\pt^W$ , $\cos\theta^{*}$
and any one of the observables defined in \eref{c6:eqn:angles}. In
fact, we perform this analysis for three different definitions of the
likelihood (L) which depend on the choice of observable, namely
$L(\pt^W,\ \cos\theta^{*},\ \cos\delta^{+} )$,
$L(\pt^W,\ \cos\theta^{*},\ \cos\delta^{-} )$ and
$L(\pt^W,\ \cos\theta^{*},\ \Delta\phi^{lV} )$.  As a first step we
produce three dimensional histograms with the various combination of
observables listed above. The choice of range and bins for each of the
observables is listed below.
\begin{itemize}
\item $\pt^W$: range $(200,1000)$~GeV, 10 bins
\item $\cos\theta^{*}$: range $(-1,1)$, 10 bins
\item $\cos\delta^{-}$: range $(-1,1)$, 10 bins
\item $\Delta\phi^{lV}$: range $(0,\pi)$, 10 bins
\end{itemize}
The histograms are binned with at least $10^4$ events after applying
all selection cuts.\\

\begin{figure}
\centering
\includegraphics[scale=0.4]{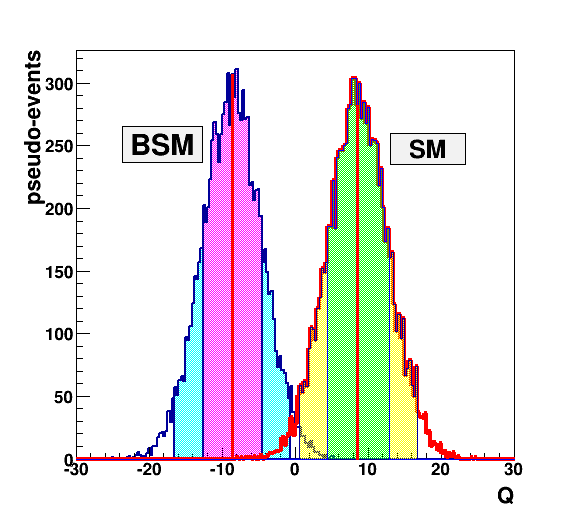}
%\scalebox{0.7}{\includegraphics{fig6_12.png}}
\caption{The distribution of $\mathcal{Q}$ from the generation of
  $1\times 10^3$ pseudo-events. For two hypotheses X=SM (right curve)
  and X=CP-odd (left curve). \label{fig:mldist}}
\end{figure}

Using these histograms we can now determine the Likelihood function.
Let $t_i$ be the expected bin height (or number of events) of the
$i^{\text{th}}$ bin derived from theory (in our case from Monte Carlo
simulations). The probability that the $i^{\text{th}}$ bin will have
$n_i$ observed events (observed bin height) is a Poissonian
probability given by
 \begin{align}
 \frac{t_i^{n_i} e^{-t_i}}{n_i!}.
 \end{align}
We can now proceed to determine the probability of generating the full
distribution for all of the histogram bins by multiplying the
probability for each of the bins. The binned likelihood is then given
by a Poisson distribution
\begin{align}
L_X=\prod_{i}^{N} \frac{t_i^{n_i} e^{-t_i}}{n_i!}.
\end{align}
Here $N$ is the number of bins, $t_i$ is the expected number of events
under the hypothesis $X$ and $n_i$ is the number of observed events.
The likelihood ratio is then defined as
\begin{align}
\mathcal{Q}=-2 Log \left(\frac{L (X | data)}{L(SM|data)} \right).
\end{align}
We now use these three dimensional histograms to generate
``pseudo-data''.  This is done by using the theoretically determined
$t_i$ to generate Poisson distributed random numbers which correspond
to our pseudo-data. We repeat this procedure for all bins in order to
generate pseudo-data. We then determine the distribution of the
likelihood ratio $\mathcal{Q}$ by generating $5 \times 10^3$
``pseudo-events''. A typical distribution for $\mathcal{Q}$ is shown
in \fref{fig:mldist}.\\

Using the distribution $\mathcal{Q}$, we can determine the p-value of
excluding the alternate (BSM hypothesis)\footnote{We use the median
  value of the null hypothesis to determine the p-value of the
  alternate hypothesis. This corresponds to twice the p-value in the
  $CL_s$ method used in LEP. }. We include the effects of backgrounds
completely but only profile over the various nuisance parameters that
arise from detector effects and selection cuts.\\

\begin{figure}
\centering
\includegraphics[scale=0.35]{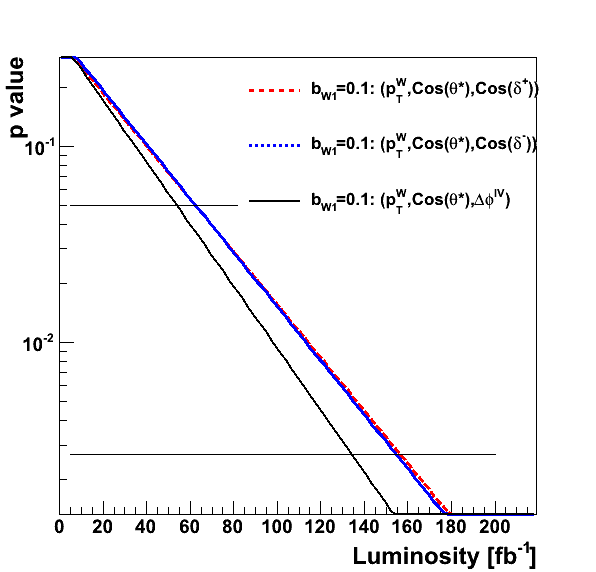}
\includegraphics[scale=0.35]{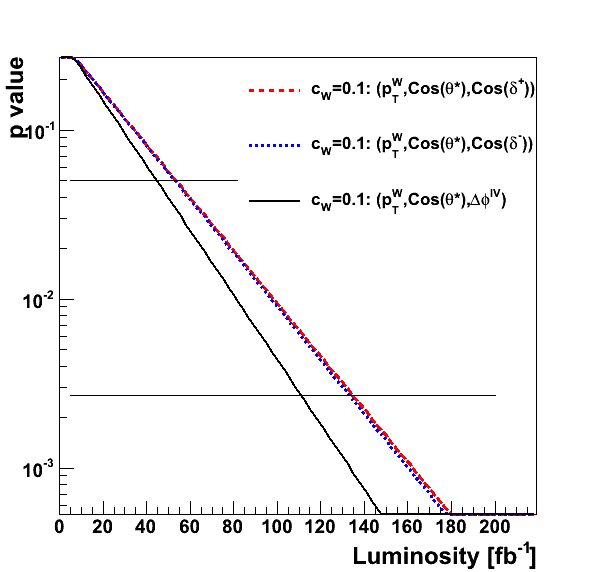}
\includegraphics[scale=0.35]{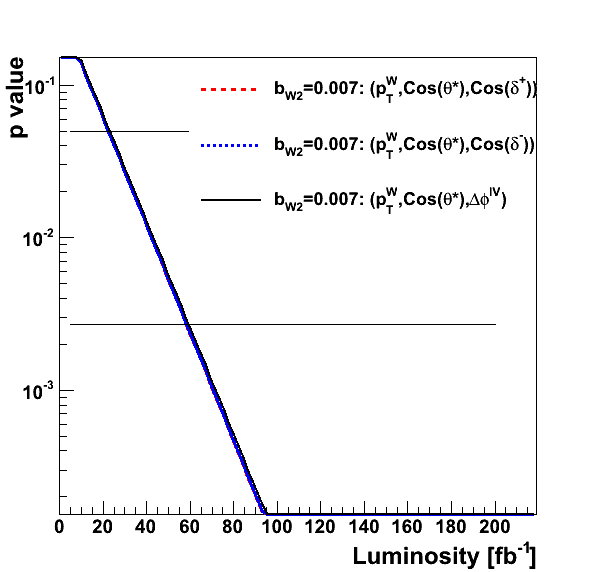}
%\scalebox{0.7}{\includegraphics{fig6_13.png}}
\caption{Plots showing the p-values for the BSM hypothesis in $Wh$
  production as a function of luminosity with three different
  likelihood functions: $L(\pt^W,\ \cos\theta^{*},\ \cos\delta^{+} )$
  (red-dashed), ($L(\pt^W,\ \cos\theta^{*},\ \cos\delta^{-} )$)(blue
  dotted) and ($L(\pt^W,\ \cos\theta^{*},\ \Delta\phi^{lV} )$) (black
  solid line). \textbf{Top left}: $\baw = 0.1$ with all other
  couplings set to zero. \textbf{Top right}: $c_W=0.1$ with all other
  couplings set to zero. \textbf{Bottom}: $\bbw=0.007$ with all other
  couplings set to zero. The coupling strengths are chosen so that
  they reproduce the SM cross-section after applying the selection
  cuts. The horizontal line on top indicates exclusion of the
  alternate hypothesis at $95\%$ confidence level. A second horizontal
  line below (if shown) indicates exclusion of the alternate
  hypothesis at $3\ \sigma$ confidence level.
\label{c6:fig:mllw1}
}
\end{figure}

The results of this procedure are shown in \fref{c6:fig:mllw1} for the
pure BSM cases, where we show the variation of the p-value of the BSM
hypothesis against the luminosity. To assess the sensitivity of each
of the observables, we set the coupling strengths to values so that
they reproduce the SM cross-section after applying all the cuts,
i.e. $(a_W=-1,\baw=0.1)$, $(a_W=-1,c_W=0.1)$ and $(a_W=-1,\bbw=0.007)$ with all other couplings
set to zero for each of these cases. This choice of couplings hence
eliminates the rate information from the analysis. We stress that this
is done to check the discriminating power of the observables under
consideration.  The horizontal line indicates exclusion of the
alternate hypothesis at $95\%$ confidence level. A second horizontal
line is shown in some cases below the first one and this indicates
exclusion of the alternate hypothesis at $3 \sigma$ confidence
level. For two of the couplings $c_W$ and $\baw$ we observe that the
likelihood constructed with the $\Delta\phi^{lV}$ observable provides
a slightly stronger discriminant. 
In both
cases we find that exclusion of the pure BSM hypothesis at $95\%$
confidence level is possible with $\sim 50 \invfb$ luminosity. The
coupling $\bbw$ can be excluded with even less data with exclusion
$95\%$ confidence level possible with just $30 \invfb$ luminosity. All
the likelihoods produce similar results in this case. This is as
expected since, the strongest discriminator for this coupling is the
transverse momentum distribution, while angular observables for $\bbw$
are not very different from SM predictions. An important point to note
is that we have set the couplings to very small values. This does not
correctly reproduce the Higgs partial decay widths. For example, if
the Higgs were a pseudo-scalar, then in order to reproduce the SM
decay width in $\dhvv$ decays, the coupling $c_V$ should have a value
$\sim 3$. For such a large value of the coupling, the $Vh$ production
channel can easily rule out the pseudo-scalar hypothesis with $\sim 20
\invfb$ of data. 
We would like to emphasize that although a pure pseudoscalar hypothesis has been ruled out by an analysis of the Higgs decaying to four lepton channel, the same is not true for the $hWW$ coupling~\cite{Chatrchyan:2013iaa}\footnote{While it is expected that the $hWW$ and $hZZ$ couplings should not be to different, from a model independent approach it is important to test both couplings independently.}. Since it is not easy to reconstruct the kinematics of the final state in the case of higgs decaying through W bosons, what we suggest is that it is possible to use only the rate information from the decay coupled with an analysis of $Wh$ production (as prescribed here) to easily rule out the pure pseudoscalar hypothesis for the $hWW$ coupling with a relatively small amount of luminosity.
\\

\begin{figure}
\centering
\includegraphics[scale=0.35]{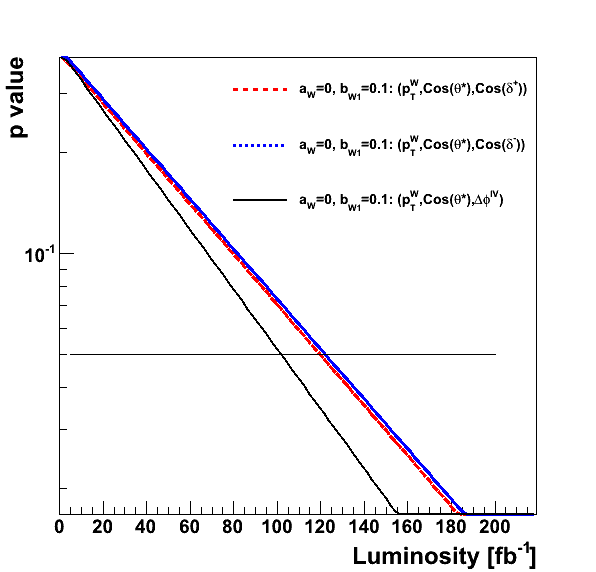}
\includegraphics[scale=0.35]{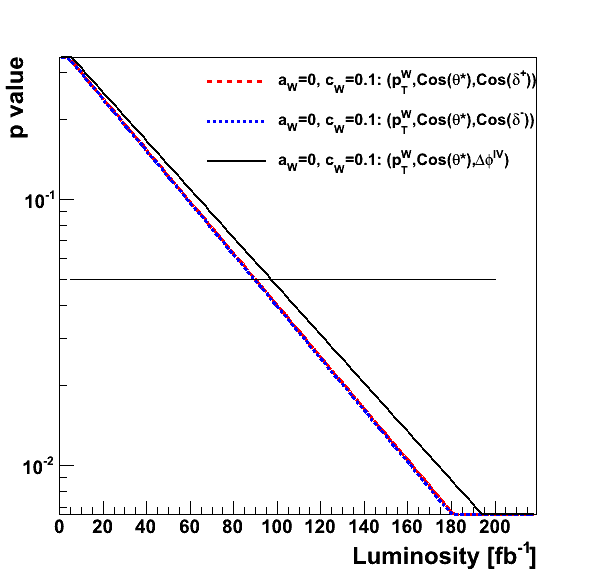}
\includegraphics[scale=0.35]{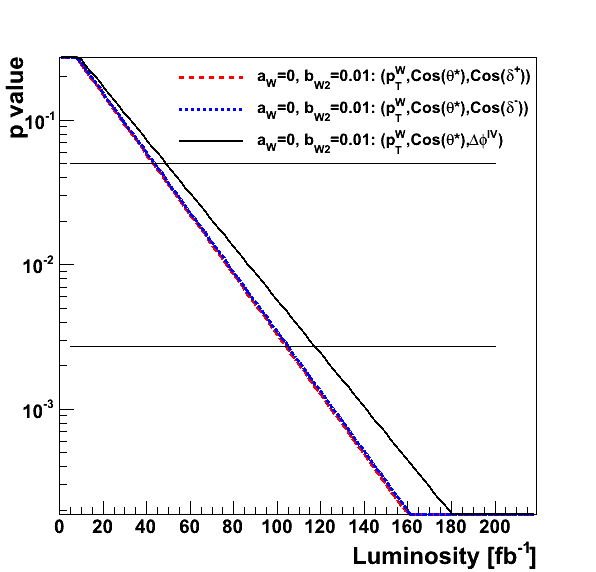}
%\scalebox{0.7}{\includegraphics{fig6_14.png}}
\caption{Plots showing the p-values for the BSM hypothesis in $Wh$
  production as a function of luminosity with three different
  likelihood functions: $L(\pt^W,\ \cos\theta^{*},\ \cos\delta^{+} )$
  (red-dashed), ($L(\pt^W,\ \cos\theta^{*},\ \cos\delta^{-} )$)(blue
  dotted) and ($L(\pt^W,\ \cos\theta^{*},\ \Delta\phi^{lV} )$) (black
  solid line). \textbf{Top left}: $(a_W=0,\baw = 0.1)$ with all other
  couplings set to zero. \textbf{Top right}: $(a_W=0,c_W=0.1)$ with
  all other couplings set to zero. \textbf{Bottom}:
  $(a_W=0,\bbw=0.01)$ with all other couplings set to zero. The
  horizontal line on top indicates exclusion of the alternate
  hypothesis at $95\%$ confidence level. A second horizontal line
  below (if shown) indicates exclusion of the alternate hypothesis at
  $3\ \sigma$ confidence level.
\label{c6:fig:mllw2}
}
\end{figure}

We are, however, more interested in the case where there are admixtures
of the SM coupling and the BSM couplings. In \fref{c6:fig:mllw2} we
show the variation of the p-value for the alternate hypothesis with
luminosity for $14$~TeV LHC.
For the case when $(a_W=0,\baw = 0.1)$, we find that the likelihood
function constructed with $\Delta\phi^{lV}$ does slightly better
than the other two likelihood functions.  We find that the BSM
hypothesis for this choice of coupling strengths can be excluded at
$95\%$ confidence level with about $100 \invfb$ luminosity. For the CP
violating case, $(a_W=0,c_W=0.1)$, as expected, the two likelihood
functions constructed with $\cos\delta^{+}$ and $\cos\delta^{-}$
appear to be the strongest discriminators with $95\%$ confidence level
exclusion of the BSM hypothesis is possible with about $100\invfb$ of
data. Finally for the case when $(a_W=0,\bbw=0.01)$, we observe, once again, that those likelihoods
constructed with $\cos\delta^{+}$ and $\cos\delta^{-}$ do slightly
better than the likelihood constructed with $\Delta\phi^{lV}$.  We find that $95\%$ confidence level exclusion of this BSM
hypothesis possible with about $50\invfb$ luminosity.\\

\begin{figure}
\centering
\includegraphics[scale=0.27]{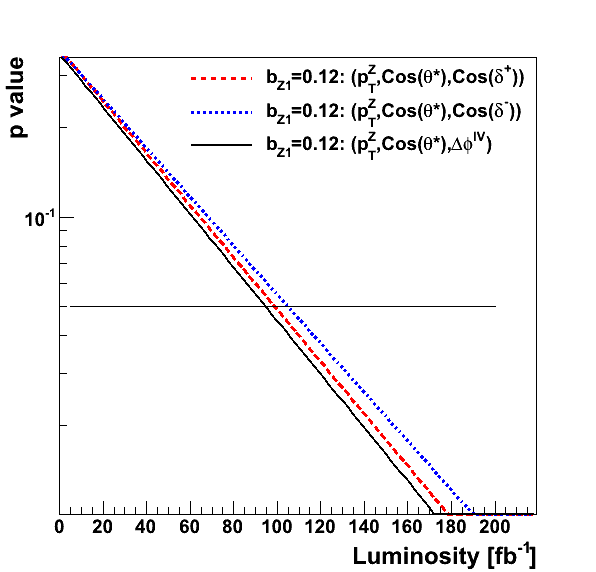}
\includegraphics[scale=0.27]{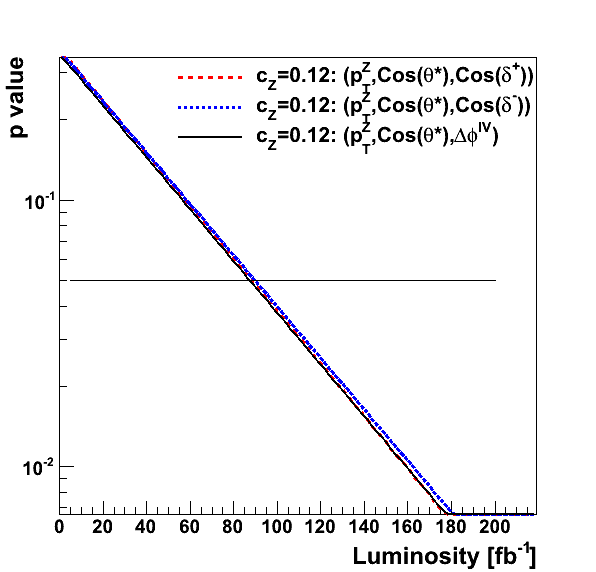}
\includegraphics[scale=0.27]{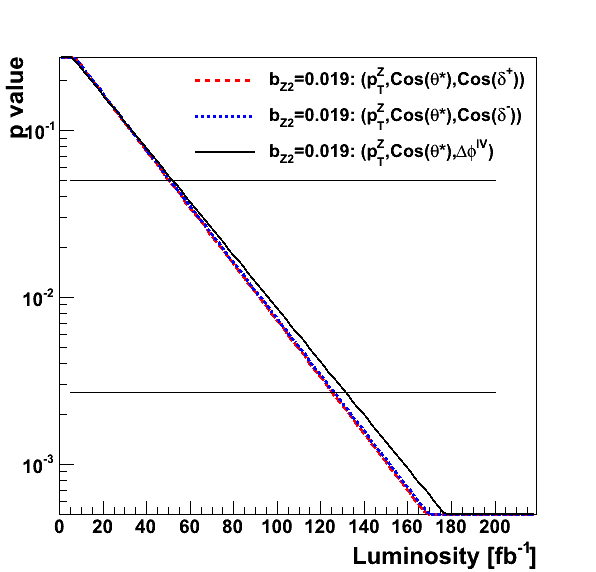}
\includegraphics[scale=0.27]{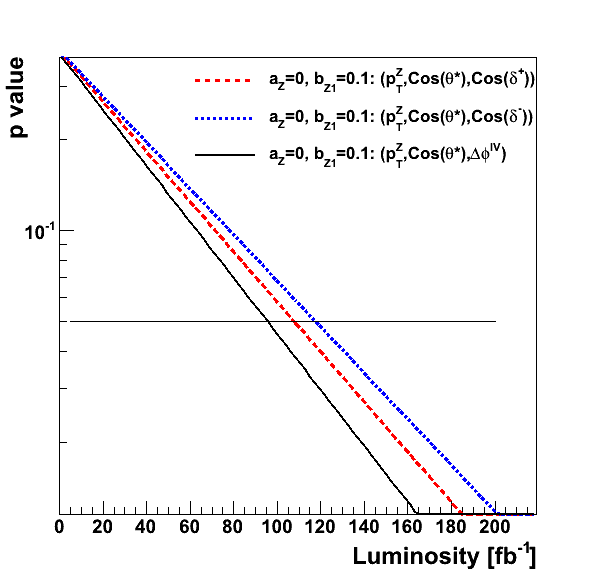}
\includegraphics[scale=0.27]{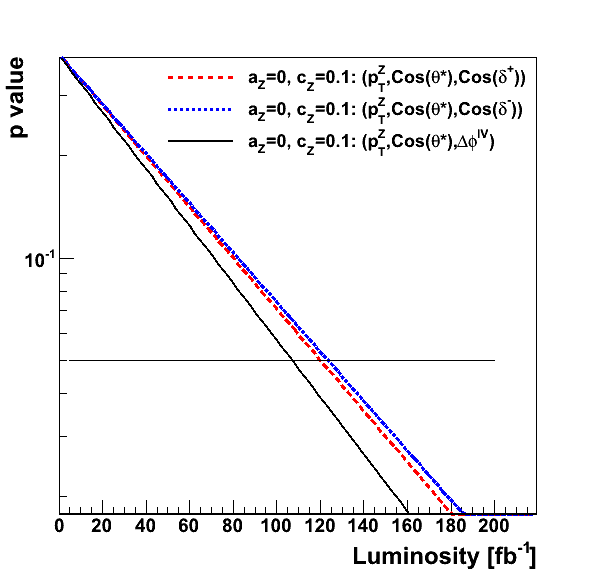}
\includegraphics[scale=0.27]{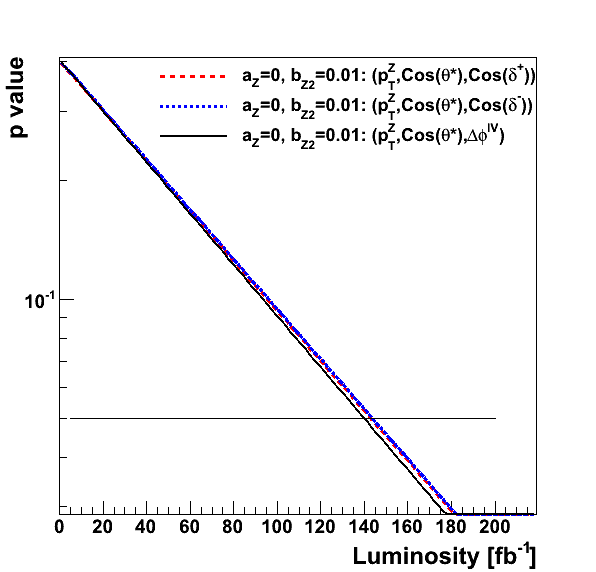}
%\scalebox{0.7}{\includegraphics{fig6_15.png}}
\caption{Plots showing the p-values for the BSM hypothesis in $Zh$
  production as a function of luminosity with three different
  likelihood functions: $L(\pt^Z,\ \cos\theta^{*},\ \cos\delta^{+} )$
  (red-dashed), ($L(\pt^Z,\ \cos\theta^{*},\ \cos\delta^{-} )$)(blue
  dotted) and ($L(\pt^Z,\ \cos\theta^{*},\ \Delta\phi^{lV} )$) (black
  solid line).  \textbf{Top row left}: $\baz = 0.12$ with all other
  couplings set to zero. \textbf{Top row right}: $c_Z=0.12$ with all
  other couplings set to zero. \textbf{Middle row left}: $\bbz=0.019$
  with all other couplings set to zero.  \textbf{Middle row left}:
  $(a_Z=0,\baz = 0.1)$ with all other couplings set to
  zero. \textbf{Bottom row left}: $(a_Z=0,c_Z=0.1)$ with all other
  couplings set to zero. \textbf{Bottom row right}:
  $(a_Z=0,\bbz=0.01)$ with all other couplings set to zero. The
  horizontal line on top indicates exclusion of the alternate
  hypothesis at $95\%$ confidence level. A second horizontal line
  below (if shown) indicates exclusion of the alternate hypothesis at
  $3\ \sigma$ confidence level.
\label{c6:fig:mllz}
}
\end{figure}

We also perform this analysis for $Zh$ production. The variation of
the p-value of the alternate hypothesis with luminosity for a $14$~TeV
LHC is shown in \fref{c6:fig:mllz}. Once again we compare the results
of three different likelihood functions constructed out of three
different combination of observables, namely
$L(\pt^Z,\ \cos\theta^{*},\ \cos\delta^{+} )$,
$L(\pt^Z,\ \cos\theta^{*},\ \cos\delta^{-} )$ and
$L(\pt^Z,\ \cos\theta^{*},\ \Delta\phi^{lV} )$. In contrast to $Wh$
production we find that all three likelihoods have a discriminating
power not very different from one another. The smaller cross-section
for $Zh$ production implies that the luminosities at which various
hypotheses can be excluded is higher than for the corresponding
hypotheses in $Wh$ production. The luminosities at which we find
exclusion of the BSM hypothesis at $95\%$ confidence level are as
follows: 
\begin{itemize}
\item $\baz=0.12$, with all other couplings set to zero : $\sim 100 \invfb$.
\item $c_Z=0.12$, with all other couplings set to zero : $\sim 90 \invfb$.
\item $\bbz=0.019$, with all other couplings set to zero : $\sim 50 \invfb$.
\item $(a_Z=0,\baz=0.1)$, with all other couplings set to zero : $\sim 100 \invfb$.
\item $(a_Z=0,c_Z=0.1)$, with all other couplings set to zero : $\sim 120 \invfb$.
\item $(a_Z=0,\bbz=0.01)$, with all other couplings set to zero : $\sim 150 \invfb$.
\end{itemize}

The luminosities listed in this section are all within the projected value of 
$300\  \invfb$ for the LHC~\cite{CMS:2013xfa}.
We reiterate here that an analysis of the $hVV$ vertex in the $Vh$
production mode was not conceived before due to the small
cross-section in this channel. We have shown that such an analysis is
indeed possible. The increased acceptance to BSM physics to the
cuts employed in a boosted analysis plays a crucial role in improving
the sensitivity of $Vh$ production to BSM physics. We have constructed observables 
that are \textit{linearly} sensitive to BSM couplings. Further,
in this section we have shown that our observables are quite
powerful and exclusion of various BSM hypotheses is possible with a
relatively small amount of data. 
The importance of our analysis is that it provides a direct method of studying the 
$hWW$ vertex. As mentioned earlier, other production modes such as VBF and $h \to W^{(*)}W^{*}$ do not provide 
clean probes of the same\footnote{For example, currently, CMS manages to exclude the pure pseudoscalar hypothesis at only $\sim 65\%$ CL using the 
$WW$ decay channel with leptonic final states~\cite{Chatrchyan:2013iaa}.}.
It should be noted that a likelihood
analysis has several nuisance parameters (from detector effects and
selection cuts) that are a source of uncertainties in this
analysis. In order to reduce the uncertainties in probing anomalous
couplings, we look at the possibility of constructing asymmetries in
the next section. This is best suited for the CP violating case where
it is easy to construct asymmetries that would vanish in case of CP
conservation. Note that it is possible to construct asymmetries that are non-zero for the SM and are also 
linear in the anomalous couplings~\footnote{For an $e^{+}e^{-}$, see for example ref.~\cite{Biswal:2005fh}}. However, here we focus our attention on CP-violation. \\

\section{Asymmetries}
\label{c6:sec:asym}

In this section, we define asymmetry parameters related to the angular
observables of section~\ref{c6:sec:ang}. There are a number of
motivations for this. Firstly, asymmetry parameters defined in terms
of ratios are typically theoretically cleaner than kinematic
distributions, due to cancellation of PDF and scale uncertainties, as
well as reduced sensitivity to radiative corrections. They are also
experimentally easier and cleaner to measure, being related to simple
counting experiments, recording the number of events in well defined
regions of phase space.\\
\begin{figure}
\centering
\includegraphics[scale=0.4]{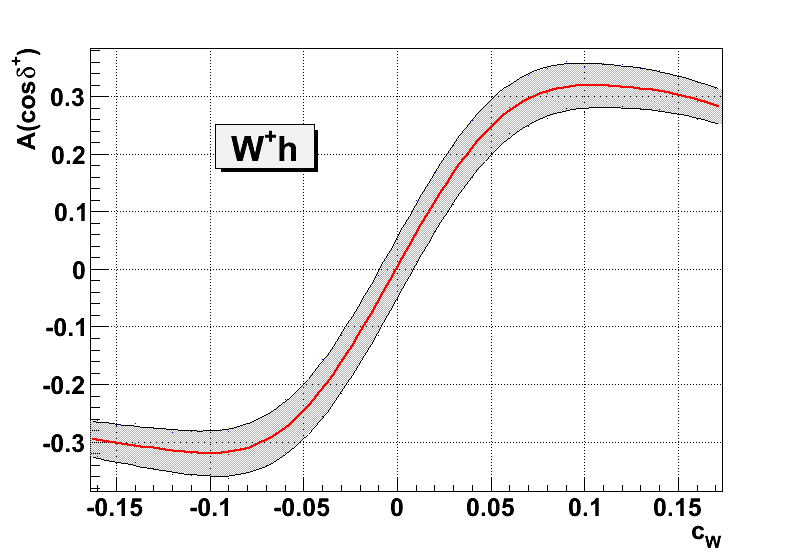}
%\scalebox{0.7}{\includegraphics{fig6_16.png}}
\caption{Variation of the asymmetry (thick red line) defined in \eref{c6:eqn:asym} with
  the strength of the coupling $c_W$. Here $a_W=0$. The shaded region denotes the $1\ \sigma$ uncertainty in the evaluation of the asymmetry using $100\  \invfb$ of data at LHC. The asymmetry and the uncertainty have been evaluated at the parton level (as described in the text) without taking into account the effect of backgrounds. 
\label{c6:fig:asym}
}
\end{figure}

Such asymmetries can be constructed using the observables
$\cos\delta^{+}$ and / or $\cos\delta^{-}$ defined in
\eref{c6:eqn:angles}. In this section we will consider the asymmetry
constructed out of the observable $\cos\delta^{+}$ only\footnote{We
  have checked that the analogous asymmetry constructed out of the
  observable $\cos\delta^{-}$ gives similarly large values for the
  asymmetry. This is expected since these two observables, as
  described earlier, are correlated.} as follows. For $W^+$ events
(tagged using the sign of the decay lepton) one defines
\begin{equation}
A(\cos\delta^{+})=\frac{\sigma(\cos\delta^{+}> 0) - \sigma(\cos\delta^{+}< 0)}{\sigma(\cos\delta^{+}> 0) + \sigma(\cos\delta^{+}< 0)},
\label{c6:eqn:asym}
\end{equation}
defining minus this quantity for $W^-$ events. For $(a_W=0,c_W=0.1)$,
we find the value of this asymmetry, after applying all selection
cuts, to be $A(\cos\delta^{+})=0.315$ for $W^{+}h$ production. We also
verify that the value of the asymmetry for all other cases (BSM, SM
and backgrounds) is less that $1\times 10^{-3}$ which is within the
statistical uncertainty limits of our procedure and can be safely
assumed to be vanishing. We emphasize that the vanishing of this
asymmetry holds true even after including detector effects. This makes
it a robust observable to probe CP violation.  Since the transverse
momentum cuts increase the acceptance of the BSM vertex, the value of
the asymmetry depends on this kinematic cut. The asymmetry also
depends on rapidity cuts since the observable depends on the
cross-product of the Higgs and gauge boson momenta.  We perform a
simple parton level analysis of $W^{+}h$ production with the $W^{+}$
decaying leptonically and the Higgs decaying to a b-quark pair. In
order to mimic the cuts of a boosted analysis, we apply the following
cuts at the parton level:
\begin{enumerate}
\item Transverse momentum of the leptons $\pt^{l}>30$~GeV; rapidity $|y^{l}|<2.5$; separation from b-quarks $\Delta R^{lb}>0.3$, where $l=\{e^{+},\mu^{+}\}$.
\item Missing transverse energy $\ptm > 30$~GeV.
\item Transverse momentum of the b-quarks $\pt^{b}>30$~GeV; rapidity $|y^{b}|<2.5$; separation between b-quarks $\Delta R^{lb}>0.1$.
\item Transverse momentum of reconstructed Higgs $\pt^{h}>200$~GeV.
\item Transverse momentum of $W^{+}$, $\pt^{W^+}>150$~GeV and $\Delta\phi(W^{+},h)> 1.2$.
\item A b-tagging efficiency of $0.6$ is used and both b-jets are tagged.
\end{enumerate}
We note that the value of the asymmetry calculated at the parton level
is in very good agreement with the asymmetry calculated using the full
boosted analysis simulation described in the previous sections
\footnote{We have only tested this for three specific values of $c_W$.}.  We
evaluate the variation of this asymmetry with the strength of the
coupling $c_W$ using a parton level analysis. The variation of this
asymmetry is shown in \fref{c6:fig:asym}. We see that the sign of the
asymmetry depends on the sign of $c_W$. We also observe that the
asymmetry peaks for a value of $c_W\sim 0.1$ while a minima is
observed for $c_W\sim -0.1$. The extrema signify regions where the
interference plays an important role. For larger values of $c_W$ the
quadratic term (in the matrix element squared) starts contributing
more strongly to the total cross-section, thus reducing the value of
this asymmetry. However, this would correspond to a kinematic region
where one no longer trusts the effective theory framework.\\

In $Zh$ production, with $(a_Z=1,c_Z=0.1)$ the asymmetry is found to
be $\sim 0.02$, an order of magnitude less than the asymmetry in case
of $Wh$ production. The much smaller value is due to the presence of
different vector and axial-vector couplings of the quarks and leptons
with the Z boson, as explained earlier.
Stronger probes of CP violation in the $hZZ$ vertex can be found in ref.~\cite{Godbole:2007cn}.
\\

\section{Conclusion}
\label{c6:sec:summary}

The ongoing attempts to pin down the nature of the recently discovered
Higgs-like particle constitute a major global effort in contemporary
particle physics. In this paper, we have considered the associated
production of the Higgs with a massive gauge boson at the LHC as an
alternative to VBF production or Higgs decays, in probing anomalous contributions to
the $hWW$ vertex. Similar analyses in VBF can be quite difficult due
to large backgrounds and our inability to reconstruct the final state
momenta. Furthermore, in VBF production there is a significant
contribution from the $hZZ$ vertex, reducing the ability to cleanly
distinguish this from the $hWW$ vertex. We have shown that such a
separation is indeed possible in $Vh$ production, despite the smaller
cross-section. This has been made possible with the use of modern
jet-substructure techniques.  Consistent with previous
studies~\cite{Djouadi:2013yb,Ellis:2012xd,Englert:2012ct}, we find
that the very same selection criteria that are applied to eliminate
backgrounds, also enhance the sensitivity to BSM physics. This is
ultimately due to the additional momentum factors that correct the
$hVV$ vertices in an effective theory framework, which boost the Higgs
to higher transverse momenta on average.\\

Building on the preliminary work of ref.~\cite{Godbole:2013saa}, we
constructed angular observables that are sensitive to new physics. To
test the ability of these observables to probe the tensor structure of
the $hVV$ vertex, we performed a log likelihood analysis. Three
dimensional likelihood functions were constructed with different
combinations of the observables. We found that with a relatively small
amount of data (less than $150\invfb$ luminosity), it is possible to
exclude all the different cases of couplings we have considered. For
example we found that the CP-violating case $(a_W=0,c_W=0.1)$ could be
excluded at $95\%$ confidence level with $\sim 90 \invfb$ luminosity
for $14$~TeV LHC.\\

Finally we constructed an asymmetry that is sensitive to the amount of
CP violation in $hWW$ interactions. The asymmetry vanishes for all
CP-conserving cases - in particular, it is zero in the SM, such that
any non-zero measurement consitutes unambiguous discovery of new
physics. We checked that the asymmetry is robust against
hadronization, radiation and detector effects. \\

The results of our paper merit further investigation, including
implementation in future experimental analyses. We furthermore
anticipate other useful applications that may result from combining
jet substructure methods with polarisation ideas. Work in this regard
is ongoing.

\section*{Acknowledgments}
DJM and CDW are supported by the UK Science and Technology Facilities
Council (STFC). 
RMG wishes to thank the
Department of Science and Technology, Government of India, for support
under grant no. SR/S2/JCB-64/2007.
DJM and CDW thank the Indian Institute of Science
for their hospitality while part of this work was carried out.

\appendix

\section{Matrix elements for $Vh$ production}
\label{c6:app}
In this appendix, we collect the matrix elements for $Vh$ production
at leading order, including the effects of the higher-dimensional
operators described in section~\ref{sec:intro}.\\

\begin{figure}
\centering
\includegraphics[scale=0.1]{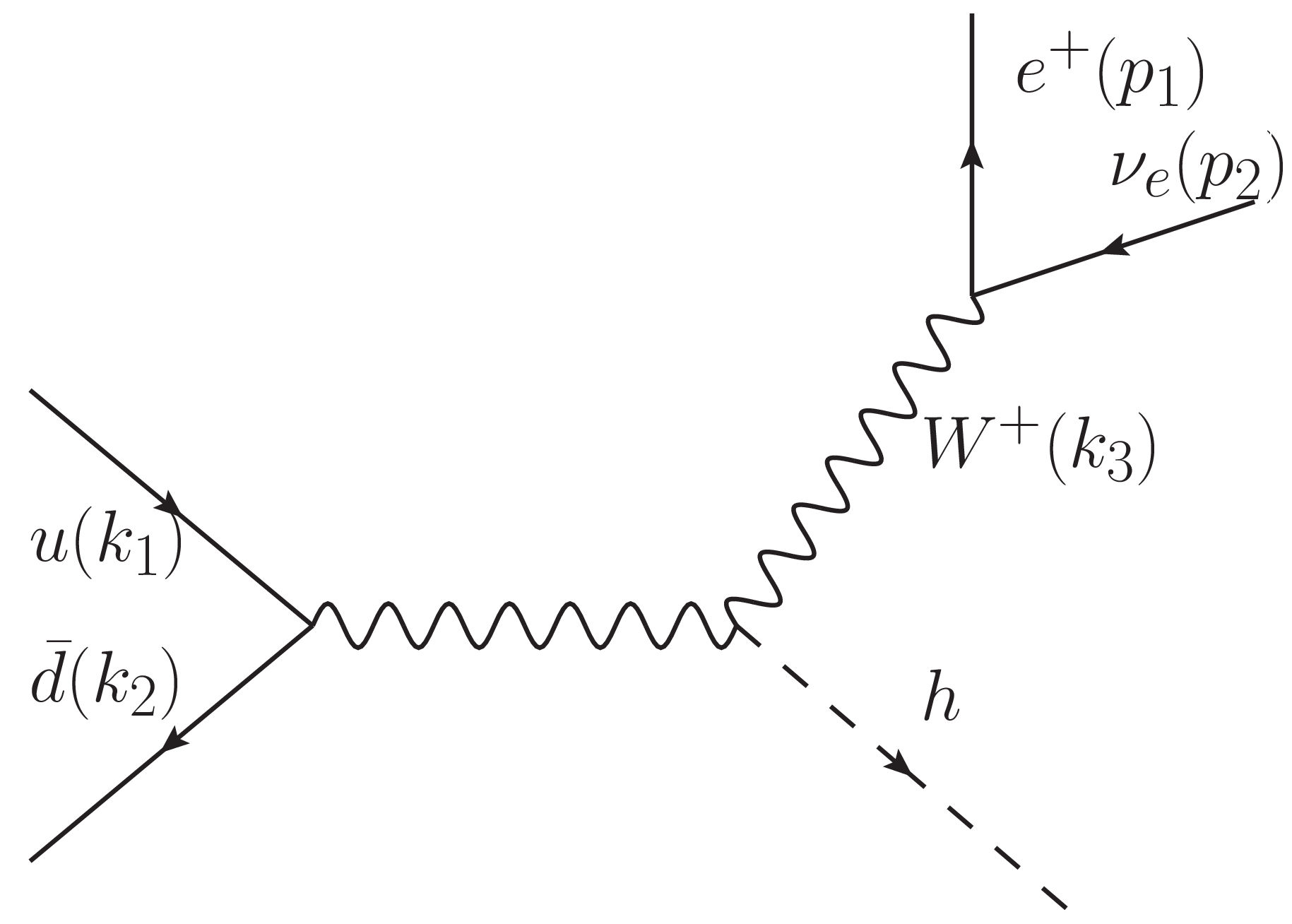}
\caption{
The Feynman diagram for the process $u\bar{d}\to W^{+} h \to e^{+} \nu_{e} h $. \label{fig:feyn-wh}
}
\end{figure}

We evaluate the squared matrix element for the Feynman diagram shown
in~\fref{fig:feyn-wh}. We do not consider decay of the Higgs boson
since we assume it to be spin zero and therefore its decay products
will not carry information about the $hWW$ vertex.  We evaluate the
matrix element squared for this process. We use the following notation
to identify parts of the matrix element that are proportional to each
of the anomalous couplings of eqs.~(\ref{c4:eqn:vertw},
\ref{c4:eqn:vertz}).
\begin{align}
|\mathcal{M}_{tot}|^2=(\mathcal{M}_{a_{W}} +\mathcal{M}_{\baw}+ \mathcal{M}_{\bbw}+ \mathcal{M}_{c_{W}})\times(\mathcal{M}_{a_{W}} +\mathcal{M}_{\baw}+ \mathcal{M}_{\bbw}+ \mathcal{M}_{c_{W}})^{\dagger},
\end{align}
where $\mathcal{M}_{i} ,\, i=\{a_{W},\baw,\bbw,c_{W}\}$ are the matrix
elements generated from the coupling with coefficient $i$.  In keeping
with the philosophy of the effective field theory approach, we keep
only terms which are at most linear in the BSM couplings (constituting
the interference of the BSM physics with the SM). Quadratic terms
would necessitate the inclusion also of dimension eight operators. The
results are:

\begin{align}
%a_{W} x a_{W}
M_{a_{W}}=\alpha_{p}|\mathcal{M}_{a_{W}}\times
\mathcal{M}^{\dagger}_{a_{W}}| &= (1+2a_W) g_{2}^6 m_{W}^2 k_1 \cdot
p_2 k_2 \cdot p_1 & \label{MaW}\\
%gha x gha
M_{a_{W}\baw}=\alpha_{p}|\mathcal{M}_{ \baw }\times \mathcal{M}^{\dagger}_{a_{W}}| &=   \left(\frac{\baw}{m_{W}^2}\right)   
g_{2}^6 m_{W}^2 ( k_1 \cdot p_2 + 
    k_2 \cdot p_1) & \nn
&( k_1 \cdot p_2 
     k_2 \cdot p_1 - 
    k_1 \cdot p_1  k_2 \cdot p_2 +
     k_1 \cdot k_2  p_1 \cdot p_2)&\\
%a_{W} x c_{W}
M_{a_{W} c_{W}}=\alpha_{p}|\mathcal{M}_{c_{W}}\times \mathcal{M}^{\dagger}_{a_{W}}|&=   \left(\frac{c_{W}}{m_{W}^2}\right) 
g_{2}^6 m_{W}^2 \epsilon^{\{k_1, k_2, p_1,p_2\}} ( k_1\cdot p_2 + 
    k_2 \cdot p_1) & \\
%a_{W} x ghb    
M_{a_{W}\bbw}=\alpha_{p}|\mathcal{M}_{a_{W}}\times \mathcal{M}^{\dagger}_{\bbw}| &=  \left(\frac{\bbw}{m_{W}^2}\right)
g_{2}^6 m_{W}^2  k_1 \cdot p_2  k_2 \cdot p_1 (p_1\cdot p_2+k_1\cdot k_2) & 
%ghb x ghb
\end{align}

Here $\epsilon^{\{k_1, k_2,
  p_1,p_2\}}=\epsilon^{\mu\nu\rho\sigma}k_{1\mu} k_{2\nu}
p_{1\rho}p_{2\sigma}$ with $\epsilon^{\mu\nu\rho\sigma}$ the
Levi-Civita tensor, and 
\begin{displaymath}
\alpha_{p}=\left(2k_1\cdot k_2
-m_{W}^2\right)^2 \left((2p_1\cdot p_2 -m_{W}^2)^2 - (m_{W}\Gamma_W)^2
\right)
\end{displaymath}
corresponds to the propagators for the W bosons. Note that the SM
contribution is included in eq.~(\ref{MaW}).

\bibliographystyle{JHEP}
\bibliography{refs.bib}

\providecommand{\href}[2]{#2}\begingroup\raggedright\begin{thebibliography}{100}

\bibitem{Aad:2012tfa}
{\bf ATLAS Collaboration} Collaboration, G.~Aad et~al., {\it {Observation of a
  new particle in the search for the Standard Model Higgs boson with the ATLAS
  detector at the LHC}},  {\em Phys.Lett.} {\bf B716} (2012) 1--29,
  [\href{http://arxiv.org/abs/1207.7214}{{\tt arXiv:1207.7214}}].

\bibitem{Chatrchyan:2012ufa}
{\bf CMS Collaboration} Collaboration, S.~Chatrchyan et~al., {\it {Observation
  of a new boson at a mass of 125 GeV with the CMS experiment at the LHC}},
  {\em Phys.Lett.} {\bf B716} (2012) 30--61,
  [\href{http://arxiv.org/abs/1207.7235}{{\tt arXiv:1207.7235}}].

\bibitem{Espinosa:2012ir}
J.~Espinosa, C.~Grojean, M.~Muhlleitner, and M.~Trott, {\it {Fingerprinting
  Higgs Suspects at the LHC}},  {\em JHEP} {\bf 1205} (2012) 097,
  [\href{http://arxiv.org/abs/1202.3697}{{\tt arXiv:1202.3697}}].

\bibitem{Ellis:2012rx}
J.~Ellis and T.~You, {\it {Global Analysis of Experimental Constraints on a
  Possible Higgs-Like Particle with Mass ~ 125 GeV}},  {\em JHEP} {\bf 1206}
  (2012) 140, [\href{http://arxiv.org/abs/1204.0464}{{\tt arXiv:1204.0464}}].

\bibitem{Ellis:2012hz}
J.~Ellis and T.~You, {\it {Global Analysis of the Higgs Candidate with Mass ~
  125 GeV}},  {\em JHEP} {\bf 1209} (2012) 123,
  [\href{http://arxiv.org/abs/1207.1693}{{\tt arXiv:1207.1693}}].

\bibitem{Espinosa:2012vu}
J.~R. Espinosa, M.~Muhlleitner, C.~Grojean, and M.~Trott, {\it {Probing for
  Invisible Higgs Decays with Global Fits}},  {\em JHEP} {\bf 1209} (2012) 126,
  [\href{http://arxiv.org/abs/1205.6790}{{\tt arXiv:1205.6790}}].

\bibitem{Corbett:2012dm}
T.~Corbett, O.~Eboli, J.~Gonzalez-Fraile, and M.~Gonzalez-Garcia, {\it
  {Constraining anomalous Higgs interactions}},  {\em Phys.Rev.} {\bf D86}
  (2012) 075013, [\href{http://arxiv.org/abs/1207.1344}{{\tt
  arXiv:1207.1344}}].

\bibitem{Espinosa:2012im}
J.~Espinosa, C.~Grojean, M.~Muhlleitner, and M.~Trott, {\it {First Glimpses at
  Higgs' face}},  {\em JHEP} {\bf 1212} (2012) 045,
  [\href{http://arxiv.org/abs/1207.1717}{{\tt arXiv:1207.1717}}].

\bibitem{Azatov:2012bz}
A.~Azatov, R.~Contino, and J.~Galloway, {\it {Model-Independent Bounds on a
  Light Higgs}},  {\em JHEP} {\bf 1204} (2012) 127,
  [\href{http://arxiv.org/abs/1202.3415}{{\tt arXiv:1202.3415}}].

\bibitem{Montull:2012ik}
M.~Montull and F.~Riva, {\it {Higgs discovery: the beginning or the end of
  natural EWSB?}},  {\em JHEP} {\bf 1211} (2012) 018,
  [\href{http://arxiv.org/abs/1207.1716}{{\tt arXiv:1207.1716}}].

\bibitem{Klute:2012pu}
M.~Klute, R.~Lafaye, T.~Plehn, M.~Rauch, and D.~Zerwas, {\it {Measuring Higgs
  Couplings from LHC Data}},  {\em Phys.Rev.Lett.} {\bf 109} (2012) 101801,
  [\href{http://arxiv.org/abs/1205.2699}{{\tt arXiv:1205.2699}}].

\bibitem{Azatov:2012ga}
A.~Azatov, R.~Contino, and J.~Galloway, {\it {Contextualizing the Higgs at the
  LHC}},  \href{http://arxiv.org/abs/1206.3171}{{\tt arXiv:1206.3171}}.

\bibitem{Gao:2010qx}
Y.~Gao, A.~V. Gritsan, Z.~Guo, K.~Melnikov, M.~Schulze, et~al., {\it {Spin
  determination of single-produced resonances at hadron colliders}},  {\em
  Phys.Rev.} {\bf D81} (2010) 075022,
  [\href{http://arxiv.org/abs/1001.3396}{{\tt arXiv:1001.3396}}].

\bibitem{Choi:2002jk}
S.~Choi, .~Miller, D.J., M.~Muhlleitner, and P.~Zerwas, {\it {Identifying the
  Higgs spin and parity in decays to Z pairs}},  {\em Phys.Lett.} {\bf B553}
  (2003) 61--71, [\href{http://arxiv.org/abs/hep-ph/0210077}{{\tt
  hep-ph/0210077}}].

\bibitem{Ellis:2012wg}
J.~Ellis and D.~S. Hwang, {\it {Does the `Higgs' have Spin Zero?}},  {\em JHEP}
  {\bf 1209} (2012) 071, [\href{http://arxiv.org/abs/1202.6660}{{\tt
  arXiv:1202.6660}}].

\bibitem{DeRujula:2010ys}
A.~De~Rujula, J.~Lykken, M.~Pierini, C.~Rogan, and M.~Spiropulu, {\it {Higgs
  look-alikes at the LHC}},  {\em Phys.Rev.} {\bf D82} (2010) 013003,
  [\href{http://arxiv.org/abs/1001.5300}{{\tt arXiv:1001.5300}}].

\bibitem{Odagiri:2002nd}
K.~Odagiri, {\it {On azimuthal spin correlations in Higgs plus jet events at
  LHC}},  {\em JHEP} {\bf 0303} (2003) 009,
  [\href{http://arxiv.org/abs/hep-ph/0212215}{{\tt hep-ph/0212215}}].

\bibitem{Buszello:2002uu}
C.~Buszello, I.~Fleck, P.~Marquard, and J.~van~der Bij, {\it {Prospective
  analysis of spin- and CP-sensitive variables in $H\rightarrow Z Z\rightarrow
  l(1)+ l(1)- l(2)+ l(2)-$ at the LHC}},  {\em Eur.Phys.J.} {\bf C32} (2004)
  209--219, [\href{http://arxiv.org/abs/hep-ph/0212396}{{\tt hep-ph/0212396}}].

\bibitem{Bredenstein:2006rh}
A.~Bredenstein, A.~Denner, S.~Dittmaier, and M.~Weber, {\it {Precise
  predictions for the Higgs-boson decay $H\rightarrow WW$ / $ZZ$ $\rightarrow$
  4 leptons}},  {\em Phys.Rev.} {\bf D74} (2006) 013004,
  [\href{http://arxiv.org/abs/hep-ph/0604011}{{\tt hep-ph/0604011}}].

\bibitem{BhupalDev:2007is}
P.~Bhupal~Dev, A.~Djouadi, R.~Godbole, M.~Muhlleitner, and S.~Rindani, {\it
  {Determining the CP properties of the Higgs boson}},  {\em Phys.Rev.Lett.}
  {\bf 100} (2008) 051801, [\href{http://arxiv.org/abs/0707.2878}{{\tt
  arXiv:0707.2878}}].

\bibitem{DeSanctis:2011yc}
U.~De~Sanctis, M.~Fabbrichesi, and A.~Tonero, {\it {Telling the spin of the
  'Higgs boson' at the LHC}},  {\em Phys.Rev.} {\bf D84} (2011) 015013,
  [\href{http://arxiv.org/abs/1103.1973}{{\tt arXiv:1103.1973}}].

\bibitem{Ellis:2012mj}
J.~Ellis, V.~Sanz, and T.~You, {\it {Prima Facie Evidence against Spin-Two
  Higgs Impostors}},  {\em Phys.Lett.} {\bf B726} (2013) 244--250,
  [\href{http://arxiv.org/abs/1211.3068}{{\tt arXiv:1211.3068}}].

\bibitem{Boughezal:2012tz}
R.~Boughezal, T.~J. LeCompte, and F.~Petriello, {\it {Single-variable
  asymmetries for measuring the `Higgs' boson spin and CP properties}},
  \href{http://arxiv.org/abs/1208.4311}{{\tt arXiv:1208.4311}}.

\bibitem{Stolarski:2012ps}
D.~Stolarski and R.~Vega-Morales, {\it {Directly Measuring the Tensor Structure
  of the Scalar Coupling to Gauge Bosons}},  {\em Phys.Rev.} {\bf D86} (2012)
  117504, [\href{http://arxiv.org/abs/1208.4840}{{\tt arXiv:1208.4840}}].

\bibitem{Ellis:2012xd}
J.~Ellis, D.~S. Hwang, V.~Sanz, and T.~You, {\it {A Fast Track towards the
  `Higgs' Spin and Parity}},  {\em JHEP} {\bf 1211} (2012) 134,
  [\href{http://arxiv.org/abs/1208.6002}{{\tt arXiv:1208.6002}}].

\bibitem{Djouadi:2013yb}
A.~Djouadi, R.~Godbole, B.~Mellado, and K.~Mohan, {\it {Probing the spin-parity
  of the Higgs boson via jet kinematics in vector boson fusion}},  {\em
  Phys.Lett.} {\bf B723} (2013) 307--313,
  [\href{http://arxiv.org/abs/1301.4965}{{\tt arXiv:1301.4965}}].

\bibitem{Godbole:2013saa}
R.~Godbole, D.~J. Miller, K.~Mohan, and C.~D. White, {\it {Boosting Higgs CP
  properties via VH Production at the Large Hadron Collider}},
  \href{http://arxiv.org/abs/1306.2573}{{\tt arXiv:1306.2573}}.

\bibitem{Ellis:2013ywa}
J.~Ellis, V.~Sanz, and T.~You, {\it {Associated Production Evidence against
  Higgs Impostors and Anomalous Couplings}},  {\em Eur.Phys.J.} {\bf C73}
  (2013) 2507, [\href{http://arxiv.org/abs/1303.0208}{{\tt arXiv:1303.0208}}].

\bibitem{Godbole:2011hw}
R.~Godbole, C.~Hangst, M.~Muhlleitner, S.~Rindani, and P.~Sharma, {\it
  {Model-independent analysis of Higgs spin and CP properties in the process
  $e^+ e^- \to t \bar t \Phi$}},  {\em Eur.Phys.J.} {\bf C71} (2011) 1681,
  [\href{http://arxiv.org/abs/1103.5404}{{\tt arXiv:1103.5404}}].

\bibitem{Muhlleitner:2012jy}
M.~Muhlleitner, R.~Godbole, C.~Hangst, S.~Rindani, and P.~Sharma, {\it
  {Analysis of Higgs spin and CP properties in a model-independent way in $e^+
  e^-\rightarrow t \bar{t} \Phi$}},  {\em Frascati Phys.Ser.} {\bf 54} (2012)
  188--197.

\bibitem{Boos:2013mqa}
E.~Boos, V.~Bunichev, M.~Dubinin, and Y.~Kurihara, {\it {Higgs boson signal at
  complete tree level in the SM extension by dimension-six operators}},
  \href{http://arxiv.org/abs/1309.5410}{{\tt arXiv:1309.5410}}.

\bibitem{Sun:2013yra}
Y.~Sun, X.-F. Wang, and D.-N. Gao, {\it {CP mixed property of the Higgs-like
  particle in the decay channel $h\to Z Z^*\to 4l$}},
  \href{http://arxiv.org/abs/1309.4171}{{\tt arXiv:1309.4171}}.

\bibitem{Einhorn:2013tja}
M.~B. Einhorn and J.~Wudka, {\it {Higgs-Boson Couplings Beyond the Standard
  Model}},  \href{http://arxiv.org/abs/1308.2255}{{\tt arXiv:1308.2255}}.

\bibitem{Anderson:2013afp}
I.~Anderson, S.~Bolognesi, F.~Caola, Y.~Gao, A.~V. Gritsan, et~al., {\it
  {Constraining anomalous HVV interactions at proton and lepton colliders}},
  \href{http://arxiv.org/abs/1309.4819}{{\tt arXiv:1309.4819}}.

\bibitem{Masso:2012eq}
E.~Masso and V.~Sanz, {\it {Limits on Anomalous Couplings of the Higgs to
  Electroweak Gauge Bosons from LEP and LHC}},  {\em Phys.Rev.} {\bf D87}
  (2013), no.~3 033001, [\href{http://arxiv.org/abs/1211.1320}{{\tt
  arXiv:1211.1320}}].

\bibitem{Nhung:2013lpa}
D.~T. Nhung, M.~Muhlleitner, J.~Streicher, and K.~Walz, {\it {Higher Order
  Corrections to the Trilinear Higgs Self-Couplings in the Real NMSSM}},  {\em
  JHEP} {\bf 1311} (2013) 181, [\href{http://arxiv.org/abs/1306.3926}{{\tt
  arXiv:1306.3926}}].

\bibitem{Heinemeyer:2013tqa}
{\bf LHC Higgs Cross Section Working Group} Collaboration, S.~Heinemeyer
  et~al., {\it {Handbook of LHC Higgs Cross Sections: 3. Higgs Properties}},
  \href{http://arxiv.org/abs/1307.1347}{{\tt arXiv:1307.1347}}.

\bibitem{Delaunay:2013npa}
C.~Delaunay, G.~Perez, H.~de~Sandes, and W.~Skiba, {\it {Higgs Up-Down CP
  Asymmetry at the LHC}},  {\em Phys.Rev.} {\bf D89} (2014) 035004,
  [\href{http://arxiv.org/abs/1308.4930}{{\tt arXiv:1308.4930}}].

\bibitem{Delaunay:2013pja}
C.~Delaunay, T.~Golling, G.~Perez, and Y.~Soreq, {\it {Charming the Higgs}},
  {\em Phys.Rev.} {\bf D89} (2014) 033014,
  [\href{http://arxiv.org/abs/1310.7029}{{\tt arXiv:1310.7029}}].

\bibitem{Maltoni:2013sma}
F.~Maltoni, K.~Mawatari, and M.~Zaro, {\it {Higgs characterisation via
  vector-boson fusion and associated production: NLO and parton-shower
  effects}},  {\em Eur.Phys.J.} {\bf 74} (2014) 2710,
  [\href{http://arxiv.org/abs/1311.1829}{{\tt arXiv:1311.1829}}].

\bibitem{Belusca-Maito:2014dpa}
H.~Belusca-Maito, {\it {Effective Higgs Lagrangian and Constraints on Higgs
  Couplings}},  \href{http://arxiv.org/abs/1404.5343}{{\tt arXiv:1404.5343}}.

\bibitem{Gavela:2014vra}
M.~Gavela, J.~Gonzalez-Fraile, M.~Gonzalez-Garcia, L.~Merlo, S.~Rigolin,
  et~al., {\it {CP violation with a dynamical Higgs}},
  \href{http://arxiv.org/abs/1406.6367}{{\tt arXiv:1406.6367}}.

\bibitem{Biekoetter:2014jwa}
A.~Biekoetter, A.~Knochel, M.~Kraemer, D.~Liu, and F.~Riva, {\it {Vices and
  Virtues of Higgs EFTs at Large Energy}},
  \href{http://arxiv.org/abs/1406.7320}{{\tt arXiv:1406.7320}}.

\bibitem{Buchmuller:1985jz}
W.~Buchmuller and D.~Wyler, {\it {Effective Lagrangian Analysis of New
  Interactions and Flavor Conservation}},  {\em Nucl.Phys.} {\bf B268} (1986)
  621.

\bibitem{Contino:2013kra}
R.~Contino, M.~Ghezzi, C.~Grojean, M.~Muhlleitner, and M.~Spira, {\it
  {Effective Lagrangian for a light Higgs-like scalar}},  {\em JHEP} {\bf 1307}
  (2013) 035, [\href{http://arxiv.org/abs/1303.3876}{{\tt arXiv:1303.3876}}].

\bibitem{Grzadkowski:2010es}
B.~Grzadkowski, M.~Iskrzynski, M.~Misiak, and J.~Rosiek, {\it {Dimension-Six
  Terms in the Standard Model Lagrangian}},  {\em JHEP} {\bf 1010} (2010) 085,
  [\href{http://arxiv.org/abs/1008.4884}{{\tt arXiv:1008.4884}}].

\bibitem{Banerjee:2013apa}
S.~Banerjee, S.~Mukhopadhyay, and B.~Mukhopadhyaya, {\it {Higher dimensional
  operators and LHC Higgs data : the role of modified kinematics}},  {\em
  Phys.Rev.} {\bf D89} (2014) 053010,
  [\href{http://arxiv.org/abs/1308.4860}{{\tt arXiv:1308.4860}}].

\bibitem{Ellis:2014dva}
J.~Ellis, V.~Sanz, and T.~You, {\it {Complete Higgs Sector Constraints on
  Dimension-6 Operators}},  {\em JHEP} {\bf 1407} (2014) 036,
  [\href{http://arxiv.org/abs/1404.3667}{{\tt arXiv:1404.3667}}].

\bibitem{Desai:2011yj}
N.~Desai, D.~K. Ghosh, and B.~Mukhopadhyaya, {\it {CP-violating HWW couplings
  at the Large Hadron Collider}},  {\em Phys.Rev.} {\bf D83} (2011) 113004,
  [\href{http://arxiv.org/abs/1104.3327}{{\tt arXiv:1104.3327}}].

\bibitem{Bolognesi:2012mm}
S.~Bolognesi, Y.~Gao, A.~V. Gritsan, K.~Melnikov, M.~Schulze, et~al., {\it {On
  the spin and parity of a single-produced resonance at the LHC}},  {\em
  Phys.Rev.} {\bf D86} (2012) 095031,
  [\href{http://arxiv.org/abs/1208.4018}{{\tt arXiv:1208.4018}}].

\bibitem{Chen:2014ona}
Y.~Chen, A.~Falkowski, I.~Low, and R.~Vega-Morales, {\it {New Observables for
  CP Violation in Higgs Decays}},  \href{http://arxiv.org/abs/1405.6723}{{\tt
  arXiv:1405.6723}}.

\bibitem{Chatrchyan:2013iaa}
{\bf CMS Collaboration} Collaboration, S.~Chatrchyan et~al., {\it {Measurement
  of Higgs boson production and properties in the WW decay channel with
  leptonic final states}},  {\em JHEP} {\bf 1401} (2014) 096,
  [\href{http://arxiv.org/abs/1312.1129}{{\tt arXiv:1312.1129}}].

\bibitem{CMS-PAS-HIG-14-014}
{\bf CMS Collaboration} Collaboration, {\it {Constraints on anomalous HVV
  interactions using H to 4l decays}},  Tech. Rep. CMS-PAS-HIG-14-014, CERN,
  Geneva, 2014.

\bibitem{2013arXiv1310.8361D}
S.~e.~a. {Dawson}, {\it {Higgs Working Group Report of the Snowmass 2013
  Community Planning Study}},  {\em ArXiv e-prints} (Oct., 2013)
  [\href{http://arxiv.org/abs/1310.8361}{{\tt arXiv:1310.8361}}].

\bibitem{ATLAS-CONF-2013-013}
{\it Measurements of the properties of the higgs-like boson in the four lepton
  decay channel with the atlas detector usi\ ng 25 fb−1 of proton-proton
  collision data},  Tech. Rep. ATLAS-CONF-2013-013, CERN, Geneva, Mar, 2013.

\bibitem{CMS-PAS-HIG-13-002}
{\it Properties of the higgs-like boson in the decay h to zz t\ o 4l in pp
  collisions at sqrt s =7 and 8 tev},  Tech. Rep. CMS-PAS-HIG-13-002, CERN,
  Geneva, 2013.

\bibitem{Chatrchyan:2012jja}
{\bf CMS Collaboration} Collaboration, S.~Chatrchyan et~al., {\it {Study of the
  Mass and Spin-Parity of the Higgs Boson Candidate Via Its Decays to Z Boson
  Pairs}},  {\em Phys.Rev.Lett.} {\bf 110} (2013) 081803,
  [\href{http://arxiv.org/abs/1212.6639}{{\tt arXiv:1212.6639}}].

\bibitem{PhysRevLett.110.081803}
{\bf CMS Collaboration} Collaboration, S.~e.~a. Chatrchyan, {\it Study of the
  mass and spin-parity of the higgs boson candidate via its decays to $z$ boson
  pairs},  {\em Phys. Rev. Lett.} {\bf 110} (Feb, 2013) 081803.

\bibitem{PhysRevD.89.092007}
{\bf (CMS Collaboration)} Collaboration, S.~e.~a. Chatrchyan, {\it Measurement
  of the properties of a higgs boson in the four-lepton final state},  {\em
  Phys. Rev. D} {\bf 89} (May, 2014) 092007.

\bibitem{Abazov:2014doa}
{\bf D0 Collaboration} Collaboration, V.~M. Abazov et~al., {\it {Constraints on
  spin and parity of the Higgs boson in $VH\rightarrow Vb\bar{b}$ final
  states}},  \href{http://arxiv.org/abs/1407.6369}{{\tt arXiv:1407.6369}}.

\bibitem{CMS-PAS-HIG-14-012}
{\bf CMS Collaboration} Collaboration, {\it {Constraints on Anomalous HWW
  Interactions using Higgs boson decays to W+W- in the fully leptonic final
  state}},  Tech. Rep. CMS-PAS-HIG-14-012, CERN, Geneva, 2014.

\bibitem{Plehn:2001nj}
T.~Plehn, D.~L. Rainwater, and D.~Zeppenfeld, {\it {Determining the structure
  of Higgs couplings at the LHC}},  {\em Phys.Rev.Lett.} {\bf 88} (2002)
  051801, [\href{http://arxiv.org/abs/hep-ph/0105325}{{\tt hep-ph/0105325}}].

\bibitem{Hankele:2006ma}
V.~Hankele, G.~Klamke, D.~Zeppenfeld, and T.~Figy, {\it {Anomalous Higgs boson
  couplings in vector boson fusion at the CERN LHC}},  {\em Phys.Rev.} {\bf
  D74} (2006) 095001, [\href{http://arxiv.org/abs/hep-ph/0609075}{{\tt
  hep-ph/0609075}}].

\bibitem{Andersen:2010zx}
J.~R. Andersen, K.~Arnold, and D.~Zeppenfeld, {\it {Azimuthal Angle
  Correlations for Higgs Boson plus Multi-Jet Events}},  {\em JHEP} {\bf 1006}
  (2010) 091, [\href{http://arxiv.org/abs/1001.3822}{{\tt arXiv:1001.3822}}].

\bibitem{Andersen:2008gc}
J.~R. Andersen, V.~Del~Duca, and C.~D. White, {\it {Higgs Boson Production in
  Association with Multiple Hard Jets}},  {\em JHEP} {\bf 0902} (2009) 015,
  [\href{http://arxiv.org/abs/0808.3696}{{\tt arXiv:0808.3696}}].

\bibitem{Miller:2001bi}
.~Miller, D.J., S.~Choi, B.~Eberle, M.~Muhlleitner, and P.~Zerwas, {\it
  {Measuring the spin of the Higgs boson}},  {\em Phys.Lett.} {\bf B505} (2001)
  149--154, [\href{http://arxiv.org/abs/hep-ph/0102023}{{\tt hep-ph/0102023}}].

\bibitem{Han:2000mi}
T.~Han and J.~Jiang, {\it {CP violating Z Z H coupling at e+ e- linear
  colliders}},  {\em Phys.Rev.} {\bf D63} (2001) 096007,
  [\href{http://arxiv.org/abs/hep-ph/0011271}{{\tt hep-ph/0011271}}].

\bibitem{Biswal:2008tg}
S.~S. Biswal, D.~Choudhury, R.~M. Godbole, and Mamta, {\it {Role of
  polarization in probing anomalous gauge interactions of the Higgs boson}},
  {\em Phys.Rev.} {\bf D79} (2009) 035012,
  [\href{http://arxiv.org/abs/0809.0202}{{\tt arXiv:0809.0202}}].

\bibitem{Biswal:2009ar}
S.~S. Biswal and R.~M. Godbole, {\it {Use of transverse beam polarization to
  probe anomalous VVH interactions at a Linear Collider}},  {\em Phys.Lett.}
  {\bf B680} (2009) 81--87, [\href{http://arxiv.org/abs/0906.5471}{{\tt
  arXiv:0906.5471}}].

\bibitem{Dutta:2008bh}
S.~Dutta, K.~Hagiwara, and Y.~Matsumoto, {\it {Measuring the Higgs-Vector boson
  Couplings at Linear $e^{+} e^{-}$ Collider}},  {\em Phys.Rev.} {\bf D78}
  (2008) 115016, [\href{http://arxiv.org/abs/0808.0477}{{\tt
  arXiv:0808.0477}}].

\bibitem{AbelleiraFernandez:2012cc}
{\bf LHeC Study Group} Collaboration, J.~Abelleira~Fernandez et~al., {\it {A
  Large Hadron Electron Collider at CERN: Report on the Physics and Design
  Concepts for Machine and Detector}},  {\em J.Phys.} {\bf G39} (2012) 075001,
  [\href{http://arxiv.org/abs/1206.2913}{{\tt arXiv:1206.2913}}].

\bibitem{Biswal:2012mp}
S.~S. Biswal, R.~M. Godbole, B.~Mellado, and S.~Raychaudhuri, {\it {Azimuthal
  Angle Probe of Anomalous $HWW$ Couplings at a High Energy $ep$ Collider}},
  {\em Phys.Rev.Lett.} {\bf 109} (2012) 261801,
  [\href{http://arxiv.org/abs/1203.6285}{{\tt arXiv:1203.6285}}].

\bibitem{Butterworth:2008iy}
J.~M. Butterworth, A.~R. Davison, M.~Rubin, and G.~P. Salam, {\it {Jet
  substructure as a new Higgs search channel at the LHC}},  {\em
  Phys.Rev.Lett.} {\bf 100} (2008) 242001,
  [\href{http://arxiv.org/abs/0802.2470}{{\tt arXiv:0802.2470}}].

\bibitem{Ellis:2009su}
S.~D. Ellis, C.~K. Vermilion, and J.~R. Walsh, {\it {Techniques for improved
  heavy particle searches with jet substructure}},  {\em Phys.Rev.} {\bf D80}
  (2009) 051501, [\href{http://arxiv.org/abs/0903.5081}{{\tt
  arXiv:0903.5081}}].

\bibitem{Ellis:2009me}
S.~D. Ellis, C.~K. Vermilion, and J.~R. Walsh, {\it {Recombination Algorithms
  and Jet Substructure: Pruning as a Tool for Heavy Particle Searches}},  {\em
  Phys.Rev.} {\bf D81} (2010) 094023,
  [\href{http://arxiv.org/abs/0912.0033}{{\tt arXiv:0912.0033}}].

\bibitem{Krohn:2009th}
D.~Krohn, J.~Thaler, and L.-T. Wang, {\it {Jet Trimming}},  {\em JHEP} {\bf
  1002} (2010) 084, [\href{http://arxiv.org/abs/0912.1342}{{\tt
  arXiv:0912.1342}}].

\bibitem{Soper:2011cr}
D.~E. Soper and M.~Spannowsky, {\it {Finding physics signals with shower
  deconstruction}},  {\em Phys.Rev.} {\bf D84} (2011) 074002,
  [\href{http://arxiv.org/abs/1102.3480}{{\tt arXiv:1102.3480}}].

\bibitem{Soper:2010xk}
D.~E. Soper and M.~Spannowsky, {\it {Combining subjet algorithms to enhance ZH
  detection at the LHC}},  {\em JHEP} {\bf 1008} (2010) 029,
  [\href{http://arxiv.org/abs/1005.0417}{{\tt arXiv:1005.0417}}].

\bibitem{Dasgupta:2013ihk}
M.~Dasgupta, A.~Fregoso, S.~Marzani, and G.~P. Salam, {\it {Towards an
  understanding of jet substructure}},  {\em JHEP} {\bf 1309} (2013) 029,
  [\href{http://arxiv.org/abs/1307.0007}{{\tt arXiv:1307.0007}}].

\bibitem{Dasgupta:2013via}
M.~Dasgupta, A.~Fregoso, S.~Marzani, and A.~Powling, {\it {Jet substructure
  with analytical methods}},  \href{http://arxiv.org/abs/1307.0013}{{\tt
  arXiv:1307.0013}}.

\bibitem{ATLAS:2012am}
{\bf ATLAS Collaboration} Collaboration, G.~Aad et~al., {\it {Jet mass and
  substructure of inclusive jets in $\sqrt{s}=7$ TeV $pp$ collisions with the
  ATLAS experiment}},  {\em JHEP} {\bf 1205} (2012) 128,
  [\href{http://arxiv.org/abs/1203.4606}{{\tt arXiv:1203.4606}}].

\bibitem{Aad:2012meb}
{\bf ATLAS Collaboration} Collaboration, G.~Aad et~al., {\it {ATLAS
  measurements of the properties of jets for boosted particle searches}},  {\em
  Phys.Rev.} {\bf D86} (2012) 072006,
  [\href{http://arxiv.org/abs/1206.5369}{{\tt arXiv:1206.5369}}].

\bibitem{Aad:2013gja}
{\bf ATLAS} Collaboration, G.~Aad et~al., {\it {Performance of jet substructure
  techniques for large-$R$ jets in proton-proton collisions at $\sqrt{s}$ = 7
  TeV using the ATLAS detector}},  {\em JHEP} {\bf 1309} (2013) 076,
  [\href{http://arxiv.org/abs/1306.4945}{{\tt arXiv:1306.4945}}].

\bibitem{Chatrchyan:2013rla}
{\bf CMS Collaboration} Collaboration, S.~Chatrchyan et~al., {\it {Studies of
  jet mass in dijet and W/Z + jet events}},  {\em JHEP} {\bf 1305} (2013) 090,
  [\href{http://arxiv.org/abs/1303.4811}{{\tt arXiv:1303.4811}}].

\bibitem{Aad:2012raa}
{\bf ATLAS Collaboration} Collaboration, G.~Aad et~al., {\it {Search for
  resonances decaying into top-quark pairs using fully hadronic decays in $pp$
  collisions with ATLAS at $\sqrt{s}=7$ TeV}},  {\em JHEP} {\bf 1301} (2013)
  116, [\href{http://arxiv.org/abs/1211.2202}{{\tt arXiv:1211.2202}}].

\bibitem{Aad:2012dpa}
{\bf ATLAS Collaboration} Collaboration, G.~Aad et~al., {\it {A search for
  $t\bar{t}$ resonances in lepton+jets events with highly boosted top quarks
  collected in $pp$ collisions at $\sqrt{s} = 7$ TeV with the ATLAS detector}},
   {\em JHEP} {\bf 1209} (2012) 041,
  [\href{http://arxiv.org/abs/1207.2409}{{\tt arXiv:1207.2409}}].

\bibitem{ATLAS:2012dp}
{\bf ATLAS Collaboration} Collaboration, G.~Aad et~al., {\it {Search for pair
  production of massive particles decaying into three quarks with the ATLAS
  detector in $\sqrt{s}=7$ TeV $pp$ collisions at the LHC}},  {\em JHEP} {\bf
  1212} (2012) 086, [\href{http://arxiv.org/abs/1210.4813}{{\tt
  arXiv:1210.4813}}].

\bibitem{ATLAS:2012ds}
{\bf ATLAS Collaboration} Collaboration, G.~Aad et~al., {\it {Search for
  pair-produced massive coloured scalars in four-jet final states with the
  ATLAS detector in proton-proton collisions at $\sqrt{s}=7$ TeV}},  {\em
  Eur.Phys.J.} {\bf C73} (2013) 2263,
  [\href{http://arxiv.org/abs/1210.4826}{{\tt arXiv:1210.4826}}].

\bibitem{Chatrchyan:2012ku}
{\bf CMS Collaboration} Collaboration, S.~Chatrchyan et~al., {\it {Search for
  anomalous t t-bar production in the highly-boosted all-hadronic final
  state}},  {\em JHEP} {\bf 1209} (2012) 029,
  [\href{http://arxiv.org/abs/1204.2488}{{\tt arXiv:1204.2488}}].

\bibitem{Chatrchyan:2012cx}
{\bf CMS Collaboration} Collaboration, S.~Chatrchyan et~al., {\it {Search for
  resonant $t\bar{t}$ production in lepton+jets events in $pp$ collisions at
  $\sqrt{s}=7$ TeV}},  {\em JHEP} {\bf 1212} (2012) 015,
  [\href{http://arxiv.org/abs/1209.4397}{{\tt arXiv:1209.4397}}].

\bibitem{Chatrchyan:2012ypy}
{\bf CMS Collaboration} Collaboration, S.~Chatrchyan et~al., {\it {Search for
  heavy resonances in the W/Z-tagged dijet mass spectrum in pp collisions at 7
  TeV}},  {\em Phys.Lett.} {\bf B723} (2013) 280--301,
  [\href{http://arxiv.org/abs/1212.1910}{{\tt arXiv:1212.1910}}].

\bibitem{Burges:1983zg}
C.~Burges and H.~J. Schnitzer, {\it {Virtual Effects of Excited Quarks as
  Probes of a Possible New Hadronic Mass Scale}},  {\em Nucl.Phys.} {\bf B228}
  (1983) 464.

\bibitem{Leung:1984ni}
C.~N. Leung, S.~Love, and S.~Rao, {\it {Low-Energy Manifestations of a New
  Interaction Scale: Operator Analysis}},  {\em Z.Phys.} {\bf C31} (1986) 433.

\bibitem{Artoisenet:2013puc}
P.~Artoisenet, P.~de~Aquino, F.~Demartin, R.~Frederix, S.~Frixione, et~al.,
  {\it {A framework for Higgs characterisation}},
  \href{http://arxiv.org/abs/1306.6464}{{\tt arXiv:1306.6464}}.

\bibitem{Alloul:2013naa}
A.~Alloul, B.~Fuks, and V.~Sanz, {\it {Phenomenology of the Higgs Effective
  Lagrangian via FeynRules}},  \href{http://arxiv.org/abs/1310.5150}{{\tt
  arXiv:1310.5150}}.

\bibitem{Alwall:2011uj}
J.~Alwall, M.~Herquet, F.~Maltoni, O.~Mattelaer, and T.~Stelzer, {\it {MadGraph
  5 : Going Beyond}},  {\em JHEP} {\bf 1106} (2011) 128,
  [\href{http://arxiv.org/abs/1106.0522}{{\tt arXiv:1106.0522}}].

\bibitem{Christensen:2008py}
N.~D. Christensen and C.~Duhr, {\it {FeynRules - Feynman rules made easy}},
  {\em Comput.Phys.Commun.} {\bf 180} (2009) 1614--1641,
  [\href{http://arxiv.org/abs/0806.4194}{{\tt arXiv:0806.4194}}].

\bibitem{Alloul:2013bka}
A.~Alloul, N.~D. Christensen, C.~Degrande, C.~Duhr, and B.~Fuks, {\it
  {FeynRules 2.0 - A complete toolbox for tree-level phenomenology}},  {\em
  Comput.Phys.Commun.} {\bf 185} (2014) 2250--2300,
  [\href{http://arxiv.org/abs/1310.1921}{{\tt arXiv:1310.1921}}].

\bibitem{Sjostrand:2006za}
T.~Sjostrand, S.~Mrenna, and P.~Z. Skands, {\it {PYTHIA 6.4 Physics and
  Manual}},  {\em JHEP} {\bf 0605} (2006) 026,
  [\href{http://arxiv.org/abs/hep-ph/0603175}{{\tt hep-ph/0603175}}].

\bibitem{Pumplin:2002vw}
J.~Pumplin, D.~Stump, J.~Huston, H.~Lai, P.~M. Nadolsky, et~al., {\it {New
  generation of parton distributions with uncertainties from global QCD
  analysis}},  {\em JHEP} {\bf 0207} (2002) 012,
  [\href{http://arxiv.org/abs/hep-ph/0201195}{{\tt hep-ph/0201195}}].

\bibitem{Cacciari:2011ma}
M.~Cacciari, G.~P. Salam, and G.~Soyez, {\it {FastJet User Manual}},  {\em
  Eur.Phys.J.} {\bf C72} (2012) 1896,
  [\href{http://arxiv.org/abs/1111.6097}{{\tt arXiv:1111.6097}}].

\bibitem{Dittmaier:2011ti}
{\bf LHC Higgs Cross Section Working Group} Collaboration, S.~Dittmaier et~al.,
  {\it {Handbook of LHC Higgs Cross Sections: 1. Inclusive Observables}},
  \href{http://arxiv.org/abs/1101.0593}{{\tt arXiv:1101.0593}}.

\bibitem{deFavereau:2013fsa}
{\bf DELPHES 3} Collaboration, J.~de~Favereau et~al., {\it {DELPHES 3, A
  modular framework for fast simulation of a generic collider experiment}},
  {\em JHEP} {\bf 1402} (2014) 057, [\href{http://arxiv.org/abs/1307.6346}{{\tt
  arXiv:1307.6346}}].

\bibitem{Han:1991ia}
T.~Han and S.~Willenbrock, {\it {QCD correction to the $p p\rightarrow W H$ and
  $Z H$ total cross-sections}},  {\em Phys.Lett.} {\bf B273} (1991) 167--172.

\bibitem{Baer:1992vx}
H.~Baer, B.~Bailey, and J.~Owens, {\it {O (alpha-s) Monte Carlo approach to W +
  Higgs associated production at hadron supercolliders}},  {\em Phys.Rev.} {\bf
  D47} (1993) 2730--2734.

\bibitem{Ohnemus:1992bd}
J.~Ohnemus and W.~J. Stirling, {\it {Order alpha-s corrections to the
  differential cross-section for the W H intermediate mass Higgs signal}},
  {\em Phys.Rev.} {\bf D47} (1993) 2722--2729.

\bibitem{Dittmaier:2012vm}
S.~Dittmaier, S.~Dittmaier, C.~Mariotti, G.~Passarino, R.~Tanaka, et~al., {\it
  {Handbook of LHC Higgs Cross Sections: 2. Differential Distributions}},
  \href{http://arxiv.org/abs/1201.3084}{{\tt arXiv:1201.3084}}.

\bibitem{Hoche:2006ph}
S.~Hoeche, F.~Krauss, N.~Lavesson, L.~Lonnblad, M.~Mangano, et~al., {\it
  {Matching parton showers and matrix elements}},
  \href{http://arxiv.org/abs/hep-ph/0602031}{{\tt hep-ph/0602031}}.

\bibitem{Englert:2012ct}
C.~Englert, M.~Spannowsky, and M.~Takeuchi, {\it {Measuring Higgs CP and
  couplings with hadronic event shapes}},  {\em JHEP} {\bf 1206} (2012) 108,
  [\href{http://arxiv.org/abs/1203.5788}{{\tt arXiv:1203.5788}}].

\bibitem{Johnson:2013zza}
{\bf D0} Collaboration, E.~Johnson, {\it {Spin and parity in the $WH
  \rightarrow \ell \nu b\bar{b}$ channel at the D0 experiment}},
  \href{http://arxiv.org/abs/1305.3675}{{\tt arXiv:1305.3675}}.

\bibitem{Han:2009ra}
T.~Han and Y.~Li, {\it {Genuine CP-odd Observables at the LHC}},  {\em
  Phys.Lett.} {\bf B683} (2010) 278--281,
  [\href{http://arxiv.org/abs/0911.2933}{{\tt arXiv:0911.2933}}].

\bibitem{Christensen:2010pf}
N.~D. Christensen, T.~Han, and Y.~Li, {\it {Testing CP Violation in ZZH
  Interactions at the LHC}},  {\em Phys.Lett.} {\bf B693} (2010) 28--35,
  [\href{http://arxiv.org/abs/1005.5393}{{\tt arXiv:1005.5393}}].

\bibitem{Englert:2012xt}
C.~Englert, D.~Goncalves-Netto, K.~Mawatari, and T.~Plehn, {\it {Higgs Quantum
  Numbers in Weak Boson Fusion}},  {\em JHEP} {\bf 1301} (2013) 148,
  [\href{http://arxiv.org/abs/1212.0843}{{\tt arXiv:1212.0843}}].

\bibitem{CMS:2013xfa}
{\bf CMS Collaboration} Collaboration, {\it {Projected Performance of an
  Upgraded CMS Detector at the LHC and HL-LHC: Contribution to the Snowmass
  Process}},  \href{http://arxiv.org/abs/1307.7135}{{\tt arXiv:1307.7135}}.

\bibitem{Biswal:2005fh}
S.~S. Biswal, R.~M. Godbole, R.~K. Singh, and D.~Choudhury, {\it {Signatures of
  anomalous VVH interactions at a linear collider}},  {\em Phys.Rev.} {\bf D73}
  (2006) 035001, [\href{http://arxiv.org/abs/hep-ph/0509070}{{\tt
  hep-ph/0509070}}].

\bibitem{Godbole:2007cn}
R.~M. Godbole, D.~Miller, and M.~M. Muhlleitner, {\it {Aspects of CP violation
  in the H ZZ coupling at the LHC}},  {\em JHEP} {\bf 0712} (2007) 031,
  [\href{http://arxiv.org/abs/0708.0458}{{\tt arXiv:0708.0458}}].

\end{thebibliography}\endgroup
\end{document}